\documentclass[prd,twocolumn,showpacs,amsmath,amssymb,nofootinbib,eqsecnum,floatfix]{revtex4}

\def\np{({\bf n}\cdot{\bf p})}
\def\pp{{\bf p}^2}
\def\ppp{({\bf p}^2)}

\newcommand{\beq}{\begin{equation}}
\newcommand{\eeq}{\end{equation}}
\newcommand{\bea}{\begin{eqnarray}} 
\newcommand{\eea}{\end{eqnarray}}
\newcommand{\ba}{\begin{array}}
\newcommand{\ea}{\end{array}}

\usepackage{graphicx}
\usepackage{bm}
\begin{document}
\title{The dynamics of precessing binary black holes using the 
post-Newtonian approximation}
\author{Michael D. Hartl}
\affiliation{
Theoretical Astrophysics, 130-33 Caltech, Pasadena CA 91125}
\author{Alessandra Buonanno}
\affiliation{groupe de Gravitation et Cosmologie (GReCO), Institut 
d'Astrophysique de Paris (CNRS), 98$^{\rm bis}$ Boulevard Arago, 
75014 Paris, France}

\begin{abstract} We investigate the (conservative) dynamics of binary black
holes using the Hamiltonian formulation of the post-Newtonian (PN) equations
of motion. The Hamiltonian we use includes spin-orbit coupling, spin-spin
coupling, and mass monopole/spin-induced quadrupole interaction terms. We
investigate the qualitative effects of these terms on the orbits; in the case
of both quasicircular and eccentric orbits, we search for the presence of
chaos (using the method of Lyapunov exponents) for a large variety of initial
conditions. For quasicircular orbits, we find
no chaotic behavior for black holes with total mass $10 \mbox{--}
40\,M_\odot$ when initially at a separation corresponding to a Newtonian
gravitational-wave (GW) frequency less than $\sim 150$~Hz. Only for rather
small initial radial distances (corresponding to a GW frequency larger than
$\sim 150$~Hz), for which spin-spin induced oscillations in the radial
separation are rather important, do we find chaotic solutions, and even then
they are rare. Moreover, these chaotic quasicircular orbits are of
questionable astrophysical significance, since they originate from direct
parametrization of the equations of motion rather than from widely separated
binaries evolving to small separations under gravitational radiation
reaction. In the case of highly eccentric orbits, which for ground-based
interferometers are not astrophysically favored, we again find chaotic
solutions, but only at pericenters so small that higher order PN corrections,
especially higher spin PN corrections, should also be taken into account.
Taken together, our surveys of quasicircular and eccentric orbits find chaos
only for orbits that are either of dubious astrophysical interest for
ground-based interferometers or which violate the approximations required for
the equations of motion to be physically valid at the post-Newtonian order
considered. \end{abstract}  \maketitle

\section{INTRODUCTION}
\label{sec:introduction_PN}

Relativistic binary systems made of compact objects, such as neutron stars or
black holes, are among the most promising (or, at least, are among the
best-understood) candidates for the production of gravitational waves
detectable by both ground- and space-based gravitational wave observatories.
The difficulty of detecting the signals from such systems has led to a
theoretical effort to understand the gravitational waveforms likely to be
emitted by such systems, which in turn has led to a consideration of their
dynamical behavior. In particular, several
authors~\cite{SuzukiMaeda1997,Levin2000,SchnittmanRasio2001,
CornishLevin2002,Hartl_2002_1} have investigated the presence of chaos in the
dynamics of compact binaries, motivated in part by the effect of such chaos on
the calculation of theoretical templates for use in matched filters. The
extreme sensitivity on initial conditions that characterizes chaotic systems
would lead to significant difficulties in the implementation of such filters,
since the number of filters would grow exponentially with increasing detection
sensitivity.

In the extreme mass-ratio limit, chaos was found in the equations describing
a \emph{spinning} particle orbiting a nonrotating (Schwarzschild) black
hole~\cite{SuzukiMaeda1997}. Refs.~\cite{Hartl_2002_1,Hartl_2002_2} extended
this result to the case of a rotating (Kerr) black hole, finding widespread
chaotic solutions. As demonstrated in~\cite{Hartl_2002_1}, however, the
values of the total spin for the test particle leading to chaotic solutions
are not realizable in physical systems. Furthermore, \cite{Hartl_2002_2}
showed that chaos, while widespread for these unrealistic spin values,
disappears in all cases for physically realistic spins. In short, there is
strong evidence that extreme mass-ratio systems, which are most relevant for
proposed space-based gravitational wave detectors, are not chaotic for any
parameter values of physical interest.

The case of the comparable-mass binaries more relevant to ground-based
gravitational-wave observatories has been investigated by several
authors~\cite{Levin2000,SchnittmanRasio2001,CornishLevin2002} using the
post-Newtonian (PN) equations of motion in the Lagrangian formalism,  using
harmonic gauge~\cite{BDIWW95,Kidder}. There was initially some doubt
regarding the results presented in~\cite{Levin2000}, which found chaos in the
PN equations for spinning bodies, since the timescale of the chaos was not
reported: it was not clear whether the chaos discovered in the equations---in
the conservative limit neglecting gravitational radiation reaction---would
have time to manifest itself in the inspiral timescale~$t_\mathrm{insp}$. 
Furthermore, the work in~\cite{SchnittmanRasio2001} cast doubt on the
presence of chaos in these systems, finding that the Lyapunov characteristic
exponents for the PN equations, which measure the divergence rate of nearby
trajectories, are zero in all cases tested. However, \cite{CornishLevin2002}
found some initial conditions, corresponding to rather eccentric orbits, 
that do have positive Lyapunov exponents, indicating the presence of chaos,
with characteristic times shorter than~$t_\mathrm{insp}$, raising the
possibility that theoretical templates calculated for systems with spinning
compact objects are affected by chaos.

In the present study, we examine and extend these results by investigating
the dynamics of spinning binary black holes using a Hamiltonian formulation
of the post-Newtonian equations of
motion~\cite{DS1988,JS1998,JS1999,DJS2000,DJSd} in the
Arnowitt-Deser-Misner (ADM) gauge. In order to make chaos formally possible, 
we exclude gravitational radiation reaction; since tests for chaos
technically require an infinite-time limit, the finite inspiral times due to
radiation reaction would eliminate the possibility of chaos. On the other
hand, we do include post-Newtonian terms involving the spin of the two
bodies: the addition of spin is essential to create the possibility of chaos,
since without spin the constants of the motion constrain the motion to be at
most quasiperiodic. As discussed in Sec.~\ref{sec:Hamiltonian}, we use
four separate spin terms in the equations of motion to model accurately their
effect of the dynamics. We focus on black holes, to the exclusion of other
compact astrophysical objects, because two of these spin terms (which involve
spin quadrupole effects) are known exactly only for black holes, and yet
their magnitudes are comparable to the spin-spin coupling and hence cannot be
ignored. (This is an extension of previous work, as other authors have not
considered these quadrupole terms when investigating chaos.)

In Sec.~\ref{sec:PN_eom} we write down the PN Hamiltonian, including spin
terms, and in Sec.~\ref{sec:PN_param} we discuss how we choose initial
conditions for quasicircular and eccentric orbits. We then investigate
chaos for comparable-mass binary black holes (Sec.~\ref{sec:PN_chaos}). Since
binary black hole inspirals tend to \emph{circularize} under gravitational
radiation reaction, we focus first on the important special case of
\emph{quasicircular} orbits and then analyze eccentric orbits. As in
previous work, we favor Lyapunov exponents (Sec.~\ref{sec:lyapunov}) to
quantify the presence (or absence) of chaos.

We work almost exclusively in geometric units ($G=c=1$). Euclidean vectors,
such as appear in the post-Newtonian equations of motion, are set in boldface,
and we use vector arrows to denote relativistic 4-vectors. The symbol $\log$
refers in all cases to the natural logarithm.

\section{The Post-Newtonian equations of motion}
\label{sec:PN_eom}

\begin{figure*}
\begin{tabular}{ccc}
\includegraphics[width=3.in]{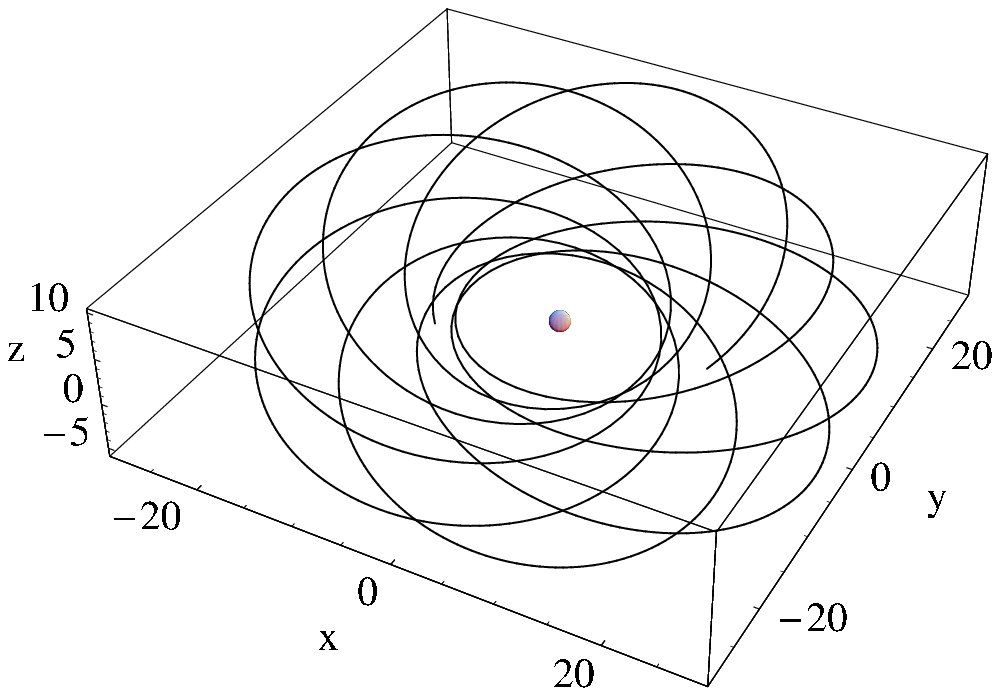} & \hspace{0.5in}
	& \includegraphics[width=2.5in]{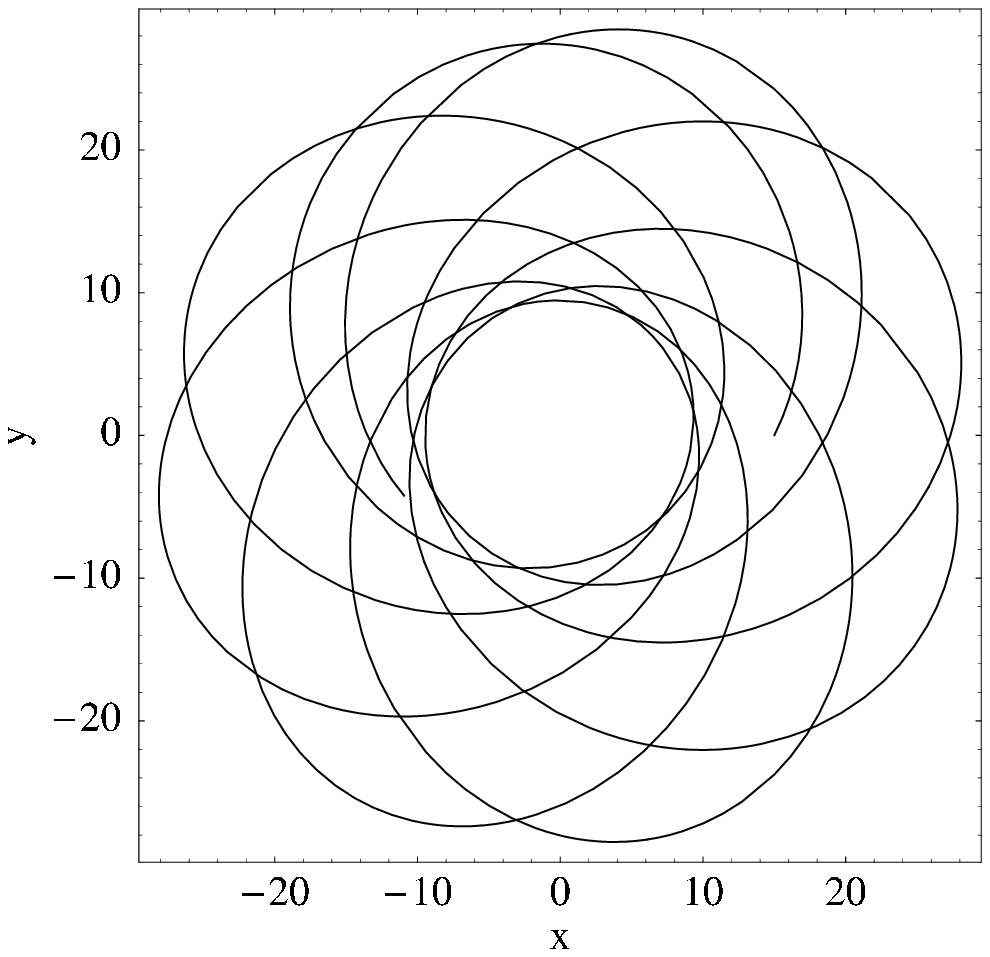}\\
(a) & & (b)\medskip\\
\end{tabular}
\caption{\label{fig:PN_eccentric_orbit}
The orbit of two maximally spinning
$10\,M_\odot$ black holes, using a Hamiltonian formulation of 
the post-Newtonian equations of motion.
(a) The orbit embedded in Euclidean space; 
(b) the projection onto the $x$-$y$ plane.
Lengths are measured in terms of the total mass $M = m_1 + m_2$, and we show a
schematic horizon at $r_\mathrm{H} = M$, indicating the collapse radius where
the relative separation of the two bodies is the sum of their horizon radii.
(We have $r_{\mathrm{H},1} = m_1$, and $r_{\mathrm{H},2} = m_2$ for
maximally spinning black
holes, so the relative separation of collapse is $r_\mathrm{H} = m_1+m_2=M$.)
Note that, in contrast to Newtonian orbits, the orbit is not closed, and the
orbital plane precesses around the center of mass.
}
\end{figure*}

The post-Newtonian (PN) equations for the two-body problem are an
approximation to full general relativity, essentially involving a series
expansion in $v/c$. As in the case of the classical two-body problem, in the
post-Newtonian case it is possible to describe the motion of a relativistic
binary in the center-of-mass frame. A typical orbit is shown in
Fig.~\ref{fig:PN_eccentric_orbit}. In this paper, we use the center-of-mass
Hamiltonian, as developed in Refs.~\cite{DS1988,JS1998,JS1999,
DJS2000,DJSd}. The Hamiltonian formulation is particularly
convenient for our present purposes: since detecting chaos involves
determining the separation of nearby \emph{phase-space} trajectories, it is
convenient to work directly in terms of spatial coordinates and their
corresponding conjugate momenta---a criterion automatically satisfied by the
Hamiltonian formulation.

\subsection{The Hamiltonian formulation}
\label{sec:Hamiltonian}

We can represent the PN Hamiltonian schematically as follows:
\begin{equation}
\label{eq:Hamiltonian}
H = H_\mathrm{N} + H_\mathrm{PN} + H_\mathrm{SO} + H_\mathrm{SS}.
\end{equation}
We include the following terms: Newtonian, post-Newtonian (usually through
2PN order, i.e., $v^4/c^4$, and sometimes through 3PN order, i.e.,
$v^6/c^6$), spin-orbit coupling (through 1.5PN order), and spin-spin coupling
(through 2PN order). (We omit the radiation reaction, as discussed in the
introduction.) Throughout this treatment, we use $\mathbf{X}$ for the
(relative) position, $\mathbf{P}$ for the conjugate (relative) momentum, and
$(\mathbf{S}_1, \mathbf{S}_2)$ for the spins of the two objects.

We denote with~$m_1$ and~$m_2$ the mass of the two bodies, and introduce the
total mass $M = m_1+m_2$ and the reduced mass $\mu = m_1 m_2/M$.\footnote{For
clarity in what follows, we adopt the arbitrary convention that $m_1\geq
m_2$.} Using these mass variables, we can express the first term (the
standard Newtonian energy) as follows:
\begin{equation}
H_\mathrm{N} = \frac{P^2}{2\mu} - \frac{\mu M}{r},
\end{equation}
where $r = |\mathbf{X}|$. The post-Newtonian terms used in this paper are
1PN and 2PN (and, in some sections, 3PN),
given by~\cite{DS1988,JS1998,JS1999,DJS2000,DJSd}:
\begin{equation}
H_\mathrm{PN} = \mu(\widehat{H}_\mathrm{1PN}+\widehat{H}_\mathrm{2PN}+ \widehat{H}_\mathrm{3PN}),
\end{equation}
where
\begin{eqnarray}
\widehat{H}_\mathrm{1PN} & = & \frac{1}{8}(3\eta-1)(\mathbf{p}^2)^2
 \nonumber \\ 
&& - \frac{1}{2}[(3+\eta)\mathbf{p}^2+
 \eta(\mathbf{n}\cdot\mathbf{p})^2]\frac{1}{q} + \frac{1}{2q^2}
\end{eqnarray}
and
\begin{eqnarray}
\widehat{H}_\mathrm{2PN} & = & \frac{1}{16}(1-5\eta+5\eta^2)(\mathbf{p}^2)^3
 \nonumber \\ 
&& + \frac{1}{8}[(5-20\eta-3\eta^2)(\mathbf{p}^2)^2
 \nonumber  \\
&&
 -2\eta^2(\mathbf{n}\cdot\mathbf{p})^2 \mathbf{p}^2
 -3\eta^4(\mathbf{n}\cdot\mathbf{p})^4]\frac{1}{q} 
 \nonumber  \\
&& 
 + \frac{1}{2}[(5+8\eta)(\mathbf{p}^2)
 + 3\eta(\mathbf{n}\cdot\mathbf{p})^2]\frac{1}{q^2}  \nonumber \\
&& - \frac{1}{4}(1+3\eta)\frac{1}{q^3}
\end{eqnarray}
and
\begin{widetext}
\bea
&& \widehat{H}_{\rm 3PN}\left({\bf q},{\bf p}\right)
= \frac{1}{128}\left(-5+35\eta-70\eta^2+35\eta^3\right)\ppp^4
\nonumber \\
&& + \frac{1}{16}\left[
\left(-7+42\eta-53\eta^2-5\eta^3\right)\ppp^3
+ (2-3\eta)\eta^2\np^2\ppp^2
+ 3(1-\eta)\eta^2\np^4\pp - 5\eta^3\np^6
\right]\frac{1}{q}
\nonumber \\
&& +\left[ \frac{1}{16}\left(-27+136\eta+109\eta^2\right)\ppp^2
+ \frac{1}{16}(17+30\eta)\eta\np^2\pp + \frac{1}{12}(5+43\eta)\eta\np^4
\right]\frac{1}{q^2} \nonumber \\
&& +\left\{ \left[ -\frac{25}{8} + \left(\frac{1}{64}\pi^2-\frac{335}{48}\right)\eta 
- \frac{23}{8}\eta^2 \right]\pp
+ \left(-\frac{85}{16}-\frac{3}{64}\pi^2-\frac{7}{4}\eta\right)\eta\np^2 
\right\}\frac{1}{q^3}
\nonumber \\
&& + \left[ \frac{1}{8} + \left(\frac{109}{12}-\frac{21}{32}\pi^2\right)\eta 
\right]\frac{1}{q^4}. \label{eq:hlast}
\eea
\end{widetext}
In the preceding formulas, we have $\eta = m_1 m_2/M^2$, and we use the unit
vector $\mathbf{n} = \mathbf{X}/r$ and the reduced canonical variables
$\mathbf{p} = \mathbf{P}/\mu$ and $\mathbf{q} = \mathbf{X}/M$. In most of
this paper, we will measure momenta in terms of $\mu$ and distances in terms
of $M$, so we will typically not distinguish between the canonical and
reduced canonical variables. In this paper, in $H_\mathrm{PN}$ we use only
the terms through the 2PN corrections (thereby omitting the 3PN terms) except
where explicitly noted.

The next term in Eq.~(\ref{eq:Hamiltonian}) is the
spin-orbit coupling (corresponding to Lense-Thirring precession in the extreme
mass-ratio limit $m_2\gg m_1$):
\begin{equation}
H_\mathrm{SO} = \frac{\mathbf{L}\cdot\mathbf{S}_\mathrm{eff}}{r^3},
\end{equation}
where
\begin{equation}
\mathbf{S}_\mathrm{eff} = 
\left(2+\frac{3}{2}\frac{m_2}{m_1}\right)\mathbf{S}_1
+ \left(2+\frac{3}{2}\frac{m_1}{m_2}\right)\mathbf{S}_2.
\end{equation}
The spin coupling term in Eq.~(\ref{eq:Hamiltonian}) has three components:
\begin{equation}
H_\mathrm{SS} = H_\mathrm{S_1S_2} + H_\mathrm{S_1S_1} + H_\mathrm{S_2S_2}.
\end{equation}
The first term, the spin-spin coupling, is
\begin{equation}
H_\mathrm{S_1S_2} =
\frac{1}{r^3}[3(\mathbf{S}_1\cdot\mathbf{n})(\mathbf{S}_2\cdot\mathbf{n})
- \mathbf{S}_1\cdot\mathbf{S}_2],
\end{equation}
which is valid for all bodies (e.g., neutron stars or white dwarfs). 
The next two terms we include are 
monopole-quadrupole
interaction
terms, and their form is valid only for black holes.\footnote{We leave
the generalization to neutron stars and other compact bodies to future work.}
They are~\cite{Poisson1998,Damour2001}
\begin{equation}
H_\mathrm{S_1S_1} = 
\frac{1}{2r^3}\,[3(\mathbf{S}_1\cdot\mathbf{n})(\mathbf{S}_1\cdot\mathbf{n})
- \mathbf{S}_1\cdot\mathbf{S}_1]\,\frac{m_2}{m_1}
\end{equation}
and
\begin{equation}
H_\mathrm{S_2S_2} = 
\frac{1}{2r^3}\,[3(\mathbf{S}_2\cdot\mathbf{n})(\mathbf{S}_2\cdot\mathbf{n})
- \mathbf{S}_2\cdot\mathbf{S}_2]\,\frac{m_1}{m_2}.
\end{equation}

With the full Hamiltonian in hand, we can now derive the equations of motion
using Poisson brackets. As in classical Hamiltonian mechanics, the
time-evolution of a dynamical 
quantity~$f(\mathbf{X}, \mathbf{P},\mathbf{S}_1,\mathbf{S}_2)$ is simply
the Poisson bracket of the quantity with the Hamiltonian:
\begin{equation}
\frac{df}{dt} = \{f, H\}.
\end{equation}
The Hamiltonian equations of motion for the (relative) position and (relative) 
momentum are then the familiar canonical equations:
\begin{equation}
\label{eq:ham_xp}
\frac{d\mathbf{X}}{dt} = +\frac{\partial H}{\partial\mathbf{P}},
\qquad
\frac{d\mathbf{P}}{dt} = -\frac{\partial H}{\partial\mathbf{X}}.
\end{equation}
To derive the spin equations of motion,
we use the canonical angular momentum Poisson bracket 
\begin{equation}
\{S^i, S^j\} = \epsilon^{ijk}S^k,
\end{equation}
which yields
\begin{equation}
\label{eq:spin1}
\frac{d\mathbf{S}_{1}}{dt} = \frac{\partial H}{\partial\mathbf{S}_{1}}
 \times \mathbf{S}_{1} \equiv \mathbf{\Omega}_1\times\mathbf{S}_{1}
\end{equation}
and
\begin{equation}
\label{eq:spin2}
\frac{d\mathbf{S}_{2}}{dt} = \frac{\partial H}{\partial\mathbf{S}_{2}}
 \times \mathbf{S}_{2}\equiv\mathbf{\Omega}_2\times\mathbf{S}_{2}.
\end{equation}
Eqs.~(\ref{eq:ham_xp}) and (\ref{eq:spin1})--(\ref{eq:spin2}), with the Hamiltonian
given by Eq.~(\ref{eq:Hamiltonian}), are the equations of
motion used throughout this paper.

\subsection{Conserved quantities}
\label{sec:conserved}

There are many conserved quantities in the post-Newtonian equations. These
constants of the motion constrain the dynamical behavior of the system and
provide valuable checks when testing a numerical implementation of the
equations. Here we discuss all the quantities known to be conserved, and at
which orders they are conserved.

\subsubsection{Quantities conserved at all orders}

The following quantities are conserved at all orders:
\begin{itemize}

\item Total energy $H$:
$\dot H = \{H, H\} = 0$ by the antisymmetry of Poisson brackets.

\item Total angular momentum $\mathbf{J} = \mathbf{L} + \mathbf{S}_1 +
\mathbf{S}_2$: see~\cite{Damour2001}.

\item The spin magnitudes $S_1$ and $S_2$: this is evident from
Eqs.~(\ref{eq:spin1}) and~(\ref{eq:spin2}), since the
cross product is perpendicular to the spin and hence can change only its
direction.

\end{itemize}

\subsubsection{Quantities conserved only through spin-orbit coupling}

If we neglect the terms quadratic in the spin (i.e., we include only terms
through spin-orbit coupling), the following additional quantities are conserved:
\begin{itemize}

\item $L^2 = \mathbf{L}\cdot\mathbf{L}$: at this order, $\dot{\mathbf{L}} =
\mathbf{S}_\mathrm{eff}\times\mathbf{L}/r^3$, which changes the
direction of $\mathbf{L}$ but not its magnitude.

\item $\mathbf{L}\cdot\mathbf{S}_\mathrm{eff}$: see~\cite{Damour2001}.

\end{itemize}
In our numerical implementation, we verify that the above quantities are
conserved at the proper orders---that is, we check that energy, total angular
momentum, and the spin magnitudes are always conserved, and verify that $L$
and $\mathbf{L}\cdot\mathbf{S}_\mathrm{eff}$ are conserved when including
only terms through spin-orbit coupling. To calculate the Lyapunov exponents
we use the techniques and routines described in Chapter 5
of~\cite{Hartl_thesis}.   For more details on our numerical implementation,
see the Appendix.

\section{Parameterizing Post-Newtonian orbits}
\label{sec:PN_param}

We discuss here two convenient methods for parameterizing initial conditions
for post-Newtonian orbits. We first describe a method that gives orbits that
approximately satisfy desired values of eccentricity, pericenter, and orbital
inclination. We then treat the important special case of quasicircular
orbits. Finally, we examine the effects of varying the post-Newtonian terms
included in the Hamiltonian.

\subsection{Eccentric orbits}
\label{sec:eccentric}

\begin{figure*}
\begin{tabular}{ccc}
\includegraphics[width=3.in]{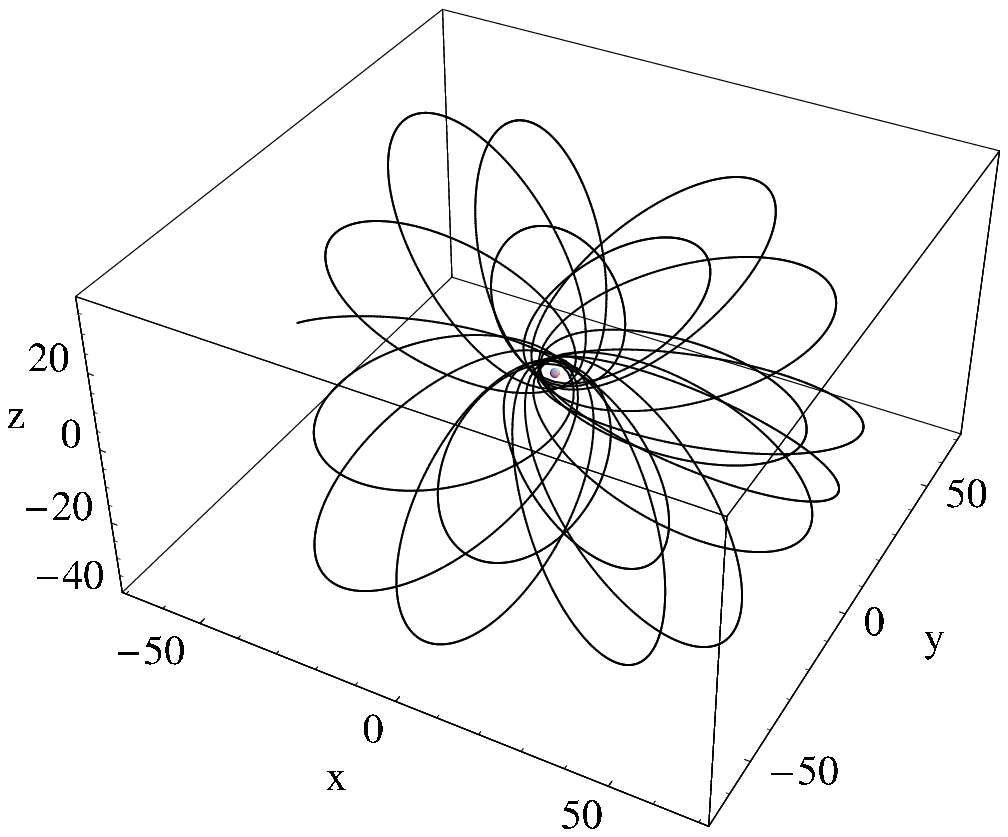} & \hspace{0.5in}
	& \includegraphics[width=2.5in]{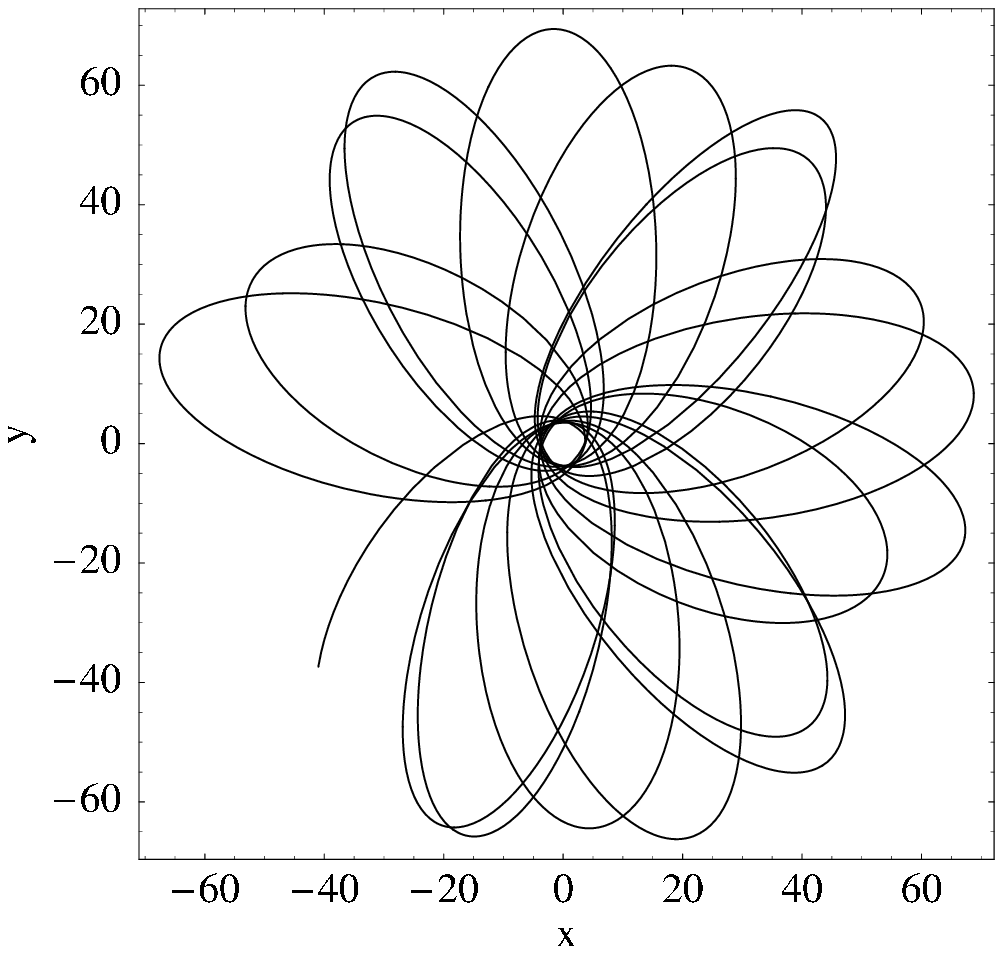}\\
(a) & & (b)\medskip\\
\end{tabular}
\caption{\label{fig:highly_eccentric_orbit}
A highly eccentric orbit for two maximally spinning $10\,M_\odot$ black
holes. (a) The orbit embedded in Euclidean space; (b) the projection onto
the $x$-$y$ plane. Lengths are measured in terms of the total mass $M = m_1 +
m_2$, and we show a schematic horizon at $r_\mathrm{H} = M$. The energy and
angular momentum of the orbit correspond to a Newtonian orbit with
eccentricity $e=0.9$, pericenter $r_p = 3.7\,M$.
The empirical values of these parameters, as determined from
the numerical solution to the equations of motion, are $e = 0.899$ and
$r_p = 3.7\,M$. The spins are $S_1 = (1/\sqrt{2}, 0, 1/\sqrt{2})$ and 
$S_2 = (1/2, 0, \sqrt{3}/2)$.
}
\end{figure*}

One convenient method for fixing initial conditions starts with the
eccentricity~$e$ and pericenter~$r_p$, which allows for the easy creation of
bound orbits and makes contact with the Newtonian limit.  For a Newtonian
orbit with given values of $(e, r_p)$, we can calculate the corresponding
(reduced) angular momentum~$L_\mathrm{N}$ and energy~$E_\mathrm{N}$ using 
\begin{equation}
L_\mathrm{N} = \sqrt{(1+e)\,r_p}
\end{equation}
and
\begin{equation}
\label{eq:E}
E_\mathrm{N} = 1-\frac{1-e}{2r_p},
\end{equation}
where we have included the rest energy in the energy term. Our
parameterization method then  involves finding a post-Newtonian orbit with
the same energy~$E$ as the Newtonian orbit, with the canonical momentum
$p_\phi$ set equal to~$L_\mathrm{N}$.\footnote{We use lower-case for $p_\phi$
(the canonical momentum conjugate to $\phi$) to distinguish it from $P_\phi =
\hat\phi\cdot\mathbf{P}$, the $\phi$ component of $\mathbf{P}$.}

The details of the eccentricity parameterization involve first setting
$\mathbf{X}_0 = (r_p, 0, 0)$ and $p_r = 0$. We then set $p_\phi =
L_\mathrm{N}/r$ (the Newtonian value of the $\phi$ momentum), leaving only
$p_\theta$ undetermined.  At this point we could set $p_\theta=0$, in
agreement with the Newtonian value; this choice leads to perfectly valid
initial conditions. But, since we are studying the dynamics of a Hamiltonian
system, we wish to assign a privileged role to the energy, and the choice
$p_\theta=0$ does not lead to a post-Newtonian system with the energy
calculated from Eq.~(\ref{eq:E}). Hence, we force the energies to agree
using 
\begin{equation}
\label{eq:HeqE}
H = E_\mathrm{Newtonian},
\end{equation}
which gives a sixth-degree polynomial equation in $p_\theta$. Since the
Newtonian value of $p_\theta$ is exactly zero, in order to produce the
post-Newtonian orbit analogous to its Newtonian counterpart we choose the
real root of Eq.~(\ref{eq:HeqE}) closest to zero. (This choice still leaves
two roots, corresponding to initial values of $p_\theta$ in the $\pm z$
direction. We arbitrarily choose the negative root, so $p_\theta$ is
initially in the $+z$ direction.)

We should note that the conditions $\mathbf{X}_0 = (r_p, 0, 0)$ and $p_r =
0$, which are chosen for computational convenience, mean that the initial
orbit has its pericenter in the $x$-$y$ plane, but this condition is not true
for a generic post-Newtonian orbit; as a result, the initial conditions do
not necessarily satisfy the requirements for a valid post-Newtonian orbit
(particularly for the nonrotating case in which orbits are confined to a
plane).  Fortunately, as soon as the orbit reaches the true pericenter---that
is, when $p_r$ is again 0---\emph{those} initial conditions do result in a
valid post-Newtonian orbit.  (This means that the radius of the pericenter
\emph{requested}---i.e., $r_p$---and the \emph{empirical} pericenter differ,
but by examining the  numerical solutions we find that the requested and
empirical pericenters typically differ by only a few percent.) Since no
finite segment of the orbit can affect the presence of chaos, which is
defined as an asymptotic property of the system (Sec.~\ref{sec:lyapunov}),
the small invalid piece at the beginning of the orbit does not affect the
final result.  

As a result of this parameterization method, we are able to find solutions to
the post-Newtonian equations of motion that have empirical values of $e$ and
$r_p$ (as determined by examining the numerical solution directly) quite
close to the values of corresponding Newtonian orbits
(Fig.~\ref{fig:highly_eccentric_orbit}). 

\subsection{Quasicircular orbits}
\label{sec:quasicircular}

A second, more specialized parameterization of the PN initial conditions
enforces the condition of \emph{quasicircularity}. In particular, through
third post-Newtonian order the quasicircular orbits are in fact exactly
circular, and even with spin-orbit coupling added there exist ``spherical''
orbits, i.e., orbits confined to lie on a sphere, with fixed radius but varying
angle~$\theta$. Once any of the spin-spin terms is turned on, exact sphericity
is impossible in general, but it is still possible to satisfy exactly the
conditions leading to spherical orbits in the absence of spin-spin coupling.
These orbits are especially important for modeling possible sources of
gravitational radiation, since the orbits of compact binaries are expected to
circularize due to gravitational radiation reaction~\cite{WillLR}.

The conditions leading to quasicircular orbits are as follows. Given an
initial radius~$r_0$, we set $\phi_0 = 0$ and $\theta_0 = \pi/2$, so that
\begin{equation}
\label{eq:quasi_X}
\mathbf{X}_0 = (r_0, 0, 0).
\end{equation}
We then require that the initial radial momentum vanish:
\begin{equation}
\label{eq:zero_Pr}
(P_r)_0 = 0.
\end{equation}
[Since the Hamiltonian is quadratic in $P_r$, this means that 
$(\dot{r})_0 =(\partial H/\partial P_r)_{P_r=0} = 0$, so that 
(at least initially) the radius
is not changing.]
Finally, we require that the initial values of $\dot{P_r}$ and $\dot{\theta}$
vanish, which means (using Hamilton's equations) that
\begin{equation}
\left(\frac{dP_r}{dt}\right)_0 = 
 -\left(\frac{\partial H}{\partial r}\right)_0 = 0
\end{equation}
and
\begin{equation}
\label{eq:quasi_Ps}
\left(\frac{d\theta}{dt}\right)_0 = 
 -\left(\frac{\partial H}{\partial P_\theta}\right)_0 = 0.
\end{equation}
Given the initial position and the initial spins, these equations can be
solved numerically for the initial values of $P_\theta$ and $P_\phi$, thereby
giving a complete set of initial conditions. 

When the spin-spin terms are included, we can always initially satisfy the
conditions for quasicircularity [Eqs.
(\ref{eq:zero_Pr})--(\ref{eq:quasi_Ps})], but these conditions are not 
preserved by the evolution, since in this case orbits with constant radial
separation no longer exist; the radial position oscillates as a function of
time. These spin-induced radial oscillations can have non-negligible
amplitudes at small separations, and it is unclear whether they represent
orbits that could result from the adiabatic inspiral of quasicircular orbits
under radiation reaction. A better method would be to set quasicircular
initial conditions when the black holes are rather far apart (so that spin
effects are negligible) and then evolve the system toward the final plunge by
including radiation reaction in the dynamics. The calculation of radiation
reaction effects for the Hamiltonian framework with spin couplings is
currently under completion~\cite{BCD}; we plan to include them in the
dynamics and investigate chaotic behaviors in the near future.

\subsection{Post-Newtonian orbits at various orders}

Here we show some of the effects of turning on or off the various
post-Newtonian terms. We use quasicircular orbits to make the effects 
especially easy to see, but this is not a necessary restriction. 
Figs.~\ref{fig:PN_eccentric_orbit} and~\ref{fig:highly_eccentric_orbit} show
that the PN equations capture the essential aspects of relativistic
orbits, such as the characteristic precession of the orbital plane.

\begin{figure*}
\begin{tabular}{ccc}
\includegraphics[width=3.in]{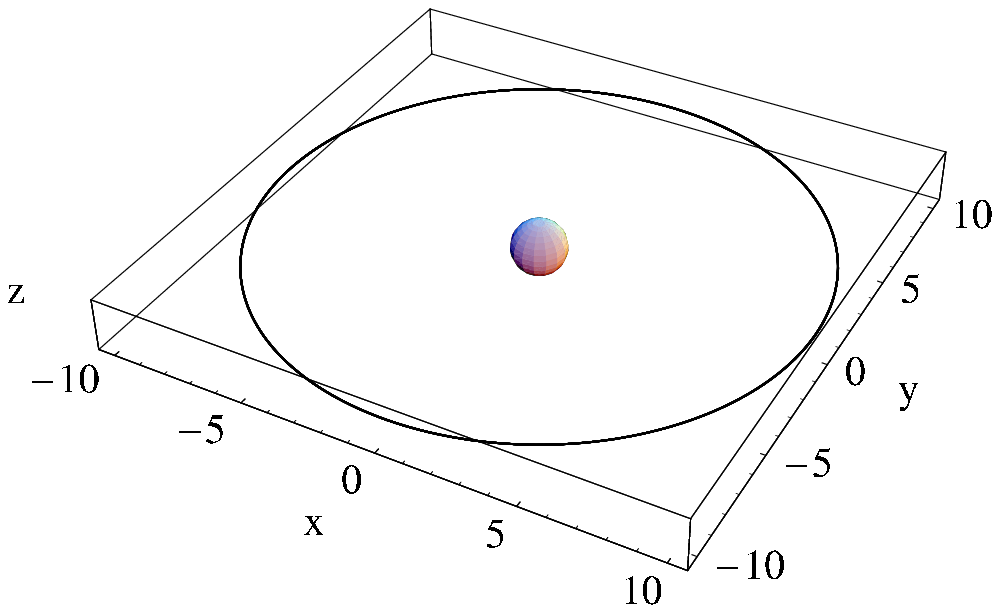} & \hspace{0.5in}
	& \includegraphics[width=2.5in]{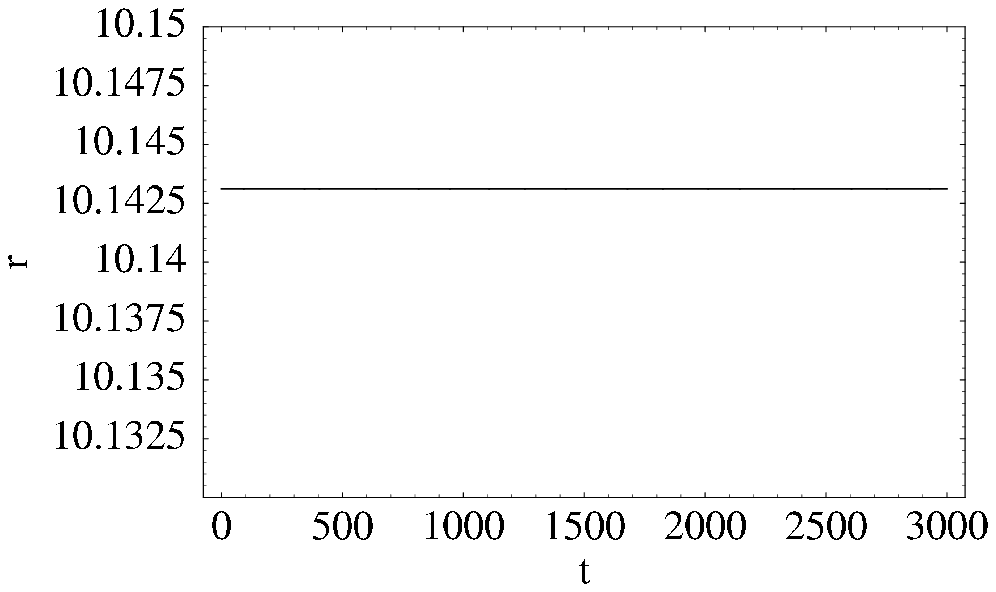}\\
(a) & & (b)\medskip\\
\end{tabular}
\caption{\label{fig:N_QC_orbit}
A post-Newtonian quasicircular orbit of two maximally spinning 
$10\,M_\odot$ black holes, with Newtonian, 1PN, and 2PN terms turned on.
(a) The orbit embedded in Euclidean space; (b) the radius $r =
\sqrt{x^2+y^2+z^2}$ as a function of time. Lengths are measured in terms of the total mass $M = m_1 + m_2$, and we
show a schematic horizon at $r_\mathrm{H} = M$.
}
\end{figure*}

\begin{figure*}
\begin{tabular}{ccc}
\includegraphics[width=3.in]{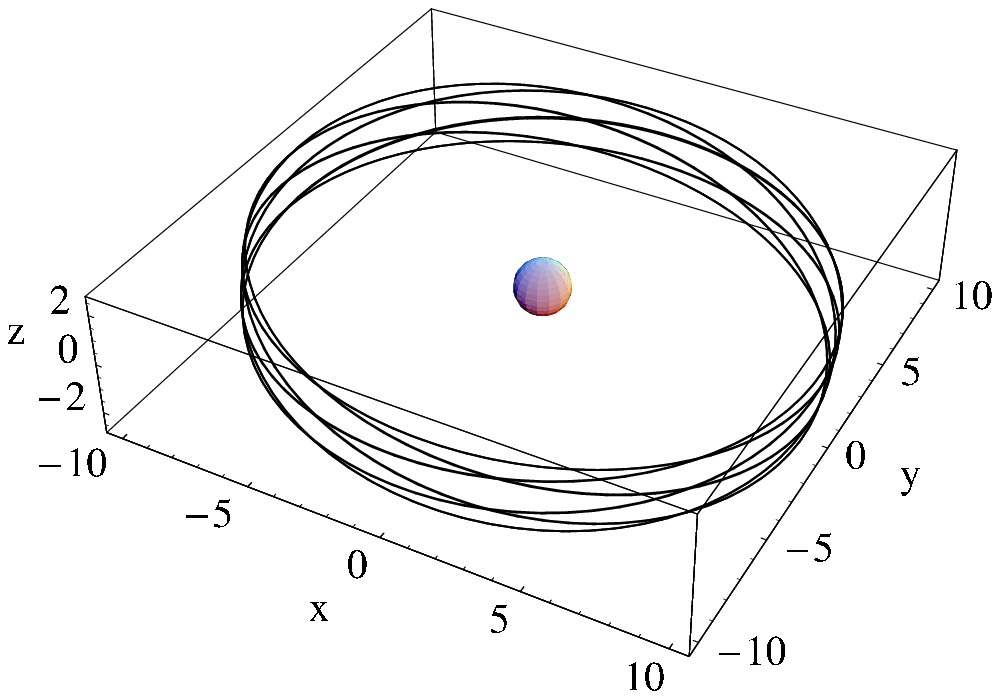} & \hspace{0.5in}
	& \includegraphics[width=2.5in]{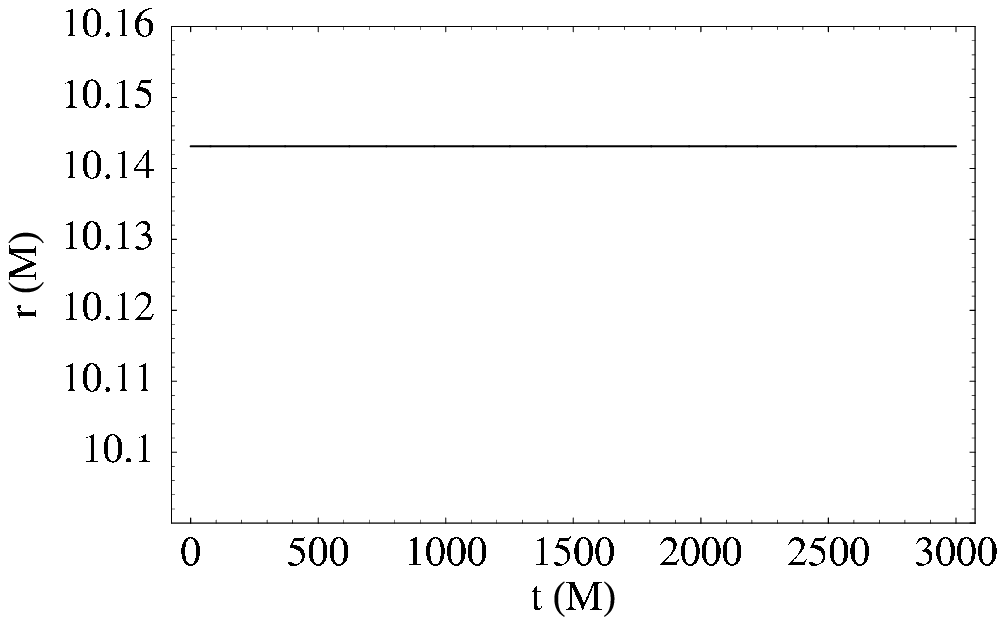}\\
(a) & & (b)\medskip\\
\end{tabular}
\caption{\label{fig:SpinOrbit_QC_orbit}
A post-Newtonian quasicircular orbit of two maximally spinning 
$10\,M_\odot$ black holes, with terms through spin-orbit coupling.
(a) The orbit embedded in Euclidean space; (b) the radius $r =
\sqrt{x^2+y^2+z^2}$ as a function of time. Lengths are measured in terms of the total mass $M = m_1 + m_2$, and we
show a schematic horizon at $r_\mathrm{H} = M$. The addition of spin-orbit
coupling to the N, 1PN, and 2PN terms destroys exact circularity, but exact
sphericity is preserved. 
}
\end{figure*}

\begin{figure*}
\begin{tabular}{ccc}
\includegraphics[width=3.in]{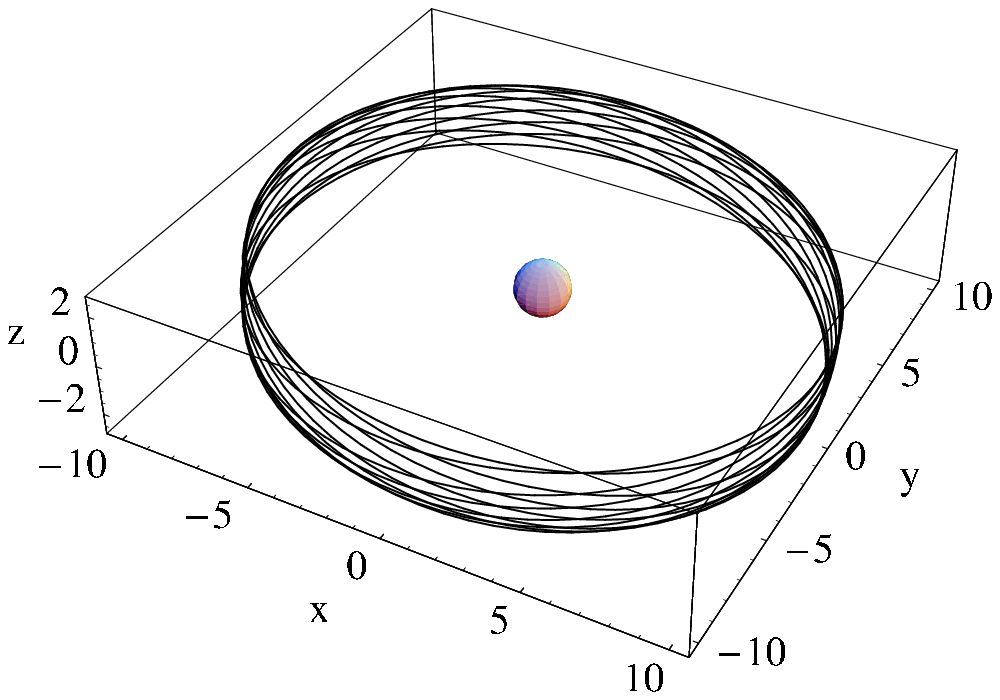} & \hspace{0.5in}
	& \includegraphics[width=2.5in]{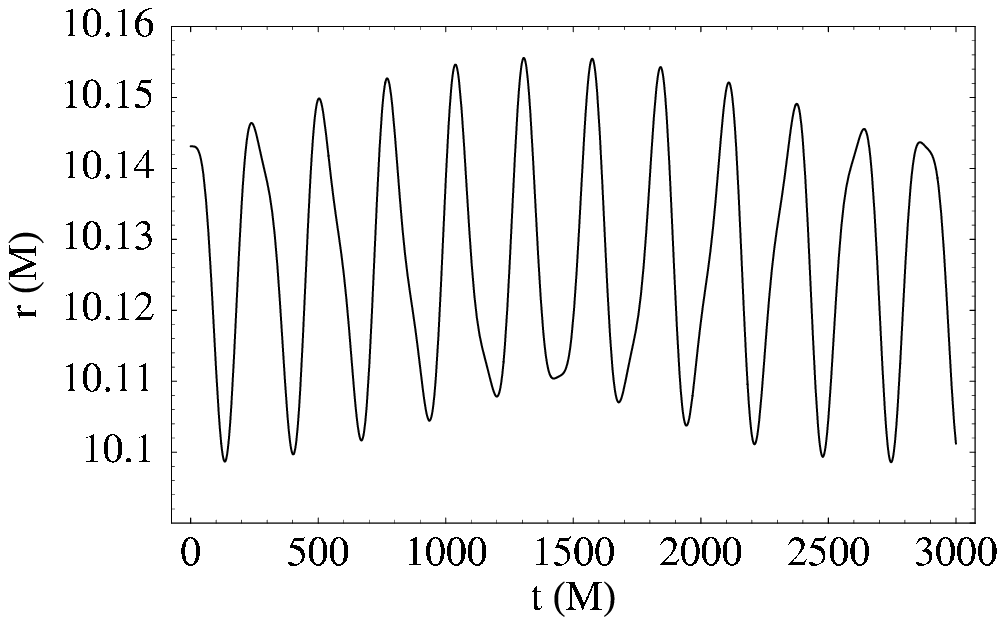}\\
(a) & & (b)\medskip\\
\end{tabular}
\caption{\label{fig:AllTerms_QC_orbit}
A post-Newtonian quasicircular orbit of two maximally spinning 
$10\,M_\odot$ black holes, with all the terms from Sec~\ref{sec:Hamiltonian}
(including the all the spin-spin couplings) present. (a) The orbit embedded
in Euclidean space; (b) the radius $r = \sqrt{x^2+y^2+z^2}$ as a function of
time. Lengths are measured in terms of the total mass $M = m_1 + m_2$, and we
show a schematic horizon at $r_\mathrm{H} = M$. The addition of spin-spin
coupling to the N, 1PN, 2PN, and spin-orbit terms destroys exact
sphericity, but Eqs.~(\ref{eq:quasi_X})--(\ref{eq:quasi_Ps}) are still
satisfied, leading to nearly circular orbits for these initial conditions.
}
\end{figure*}

Figs.~\ref{fig:N_QC_orbit}--\ref{fig:AllTerms_QC_orbit} show quasicircular
orbits satisfying Eqs.~(\ref{eq:quasi_X})--(\ref{eq:quasi_Ps}), with increasing
orders (up to 2PN) of the PN Hamiltonian included. The third figure includes all the
spin-spin terms ($S_1S_2$, $S_1S_1$, and $S_2S_2$); we do not show any orbits
with subsets of these terms activated because, to the precision visible in the
figure, the $S_1S_2$ term dominates, and the figure is identical with either of
the other terms removed.

\section{Investigating chaos in the Post-Newtonian equations}
\label{sec:PN_chaos}

\begin{figure}
\begin{center}
\includegraphics[width=3in]{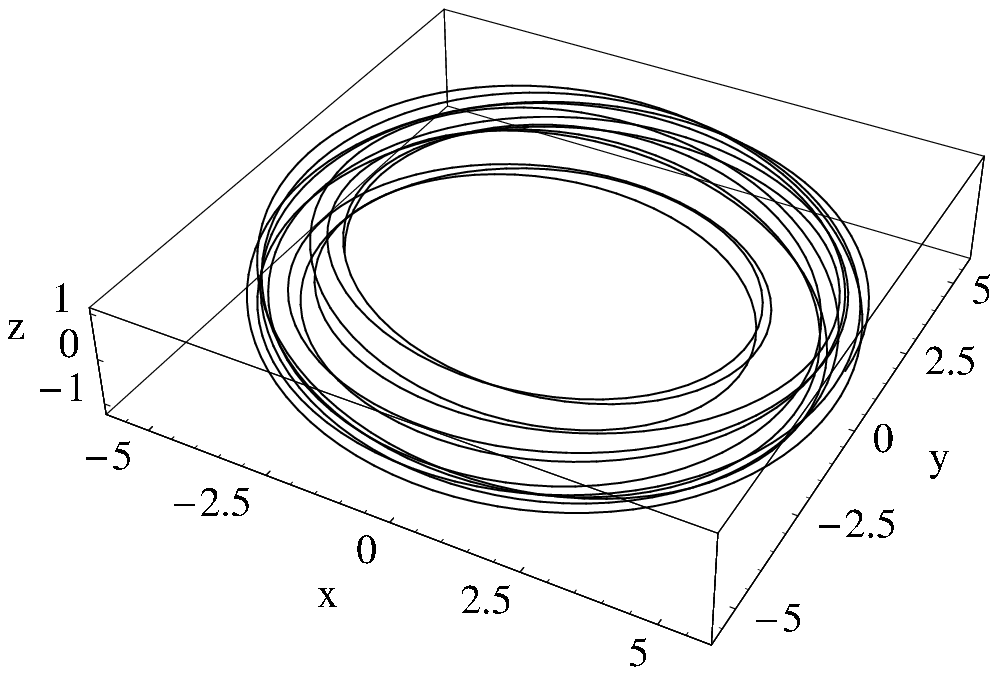}
\end{center}
\caption[Chaotic orbit of two maximally spinning $10\,M_\odot$ black holes]{
\label{fig:PN_orbit}
The orbit of two maximally spinning $10\,M_\odot$ black holes. The dynamics
are chaotic, as shown in Fig.~\ref{fig:PN_Lyapunov}. The initial conditions
satisfy the requirements for quasicircularity, though in fact the orbit's
radius is not even approximately constant (see
Sec.~\ref{sec:varying_frequencies} below). The initial radius 
is $5.658\,M$, corresponding to an orbit
with a Newtonian gravitational-wave frequency of
$f_\mathrm{GW}=240~\textrm{Hz}$ [Eq.~(\ref{eq:fGW})]. 
The initial spins are $\mathbf{S}_1 = (0.13036, 0.262852, -0.955989)\,m_1^2$ 
and $\mathbf{S}_2 = 
(0.118966, -0.13459, -0.983734)\,m_2^2$.
The other initial conditions are fixed by the conditions for quasicircularity
(Sec.~\ref{sec:quasicircular}).
}
\end{figure}

\begin{figure}
\begin{center}
\includegraphics[width=3in]{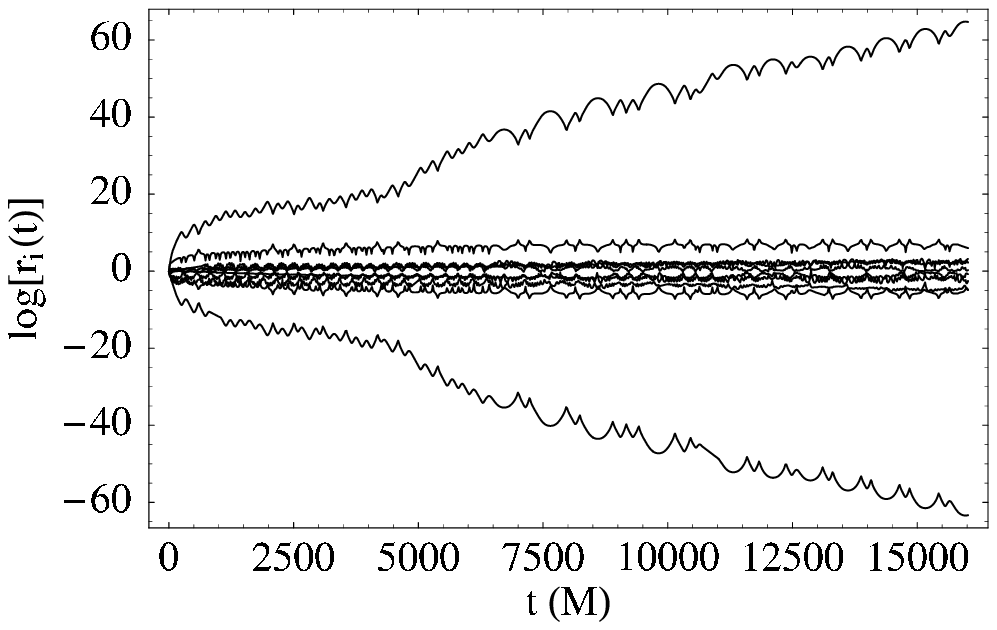}
\end{center}
\caption{
\label{fig:PN_Lyapunov}
The natural logarithms of the ellipsoid 
axes~$r_i$ vs.~$t$ for the system shown in Fig.~\ref{fig:PN_orbit}. The slopes
of the lines are the Lyapunov exponents. 
Two nonzero exponents are clearly visible 
($\lambda =\pm3.2\times10^{-3}\,M^{-1}$), but all the others are consistent
with zero. The curve with the largest slope corresponds to the upper line in
Fig.~\ref{fig:lyapunov_nonzero}.
There is an apparent~$\pm\lambda$ symmetry: for each exponent
$+\lambda$, there is a corresponding exponent~$-\lambda$; even the zero
exponents approach zero symmetrically. This behavior is a characteristic of 
Hamiltonian systems~\cite{EckmannRuelle1985}.}
\end{figure}

Previous studies of chaos in the post-Newtonian equations considered
comparable mass-ratio binaries with eccentric
orbits~\cite{LevinPRL2000,Levin2000,CornishLevin2002,
CornishLevinSR2002,CornishLevin1997}. Here, we also consider comparable
mass-ratio binaries and focus first on quasicircular orbits and 
then on eccentric orbits. As noted in the introduction, these orbits are particularly important because many
astrophysically relevant binary systems should circularize due to the energy
lost to gravitational radiation. We set the radii of these orbits so that
the frequencies of their corresponding gravitational waves lie in a range
$40~\textrm{Hz} < f_\mathrm{GW} < 240~\textrm{Hz}$, roughly corresponding to
the frequency band for the Laser Interferometer Gravitational-Wave
Observatory (LIGO) and VIRGO. [More properly, we choose radii such that the
\emph{Newtonian} frequency of the gravitational waves (which is simply twice
the orbital frequency) lies in the LIGO/VIRGO band, as discussed in
Sec.~\ref{sec:quasicircular_survey} below.] For initial separations such 
that the Newtonian GW frequency is smaller than 40 Hz, we never find chaotic 
quasicircular orbits.

\subsection{A brief discussion of Lyapunov exponents}
\label{sec:lyapunov}

As in previous works~\cite{Hartl_2002_1,Hartl_2002_2}, Lyapunov exponents are
our primary tool for investigating the nonlinear dynamics of general
relativistic systems. We have discussed at length
in~\cite{Hartl_2002_1,Hartl_thesis} our techniques for calculating these
exponents for systems similar to the PN equations. Here we present a brief
summary of Lyapunov exponents.

Given an initial condition in the phase space of a dynamical system with
$n$~degrees of freedom, we imagine an $n$-dimensional ball of nearby initial
conditions centered on that point. As the dynamics unfold, in general the
ball is stretched in some directions and squeezed in others, deforming into
an $n$-dimensional ellipsoid under the action of the flow. Such an ellipsoid
has~$n$ principal (semi)axes, and the average rate of stretching or squeezing
of each axis is a \emph{Lyapunov number}, whose natural logarithm is the
\emph{Lyapunov exponent} associated with the axis. In general, the $i$th
Lyapunov exponent of a dynamical system is 
\begin{equation}
\label{eq:lyap}
\lambda_i = \lim_{t\rightarrow\infty}\frac{\log{[r_i(t)]}}{t},
\end{equation}
where $r_i(t)$ is the $i$th principal ellipsoid axis. Implementing this
prescription numerically leads to a visualization of the exponents as a plot
of $\log{[r_i(t)]}$ vs.~$t$ (so that the slope is the exponent $\lambda_i$),
which we refer to as a \emph{Lyapunov plot}. The result for a chaotic PN
orbit appears in Figs.~\ref{fig:PN_orbit} and~\ref{fig:PN_Lyapunov}.

In practice, following the evolution of the phase-space ellipsoid, and
thereby extracting all the Lyapunov exponents of the system, involves using
the Jacobian matrix of the system to model an ``infinitesimal'' ball that
captures the true linear approximation to the dynamics. It is also possible,
and computationally faster, to extract only the largest Lyapunov exponent by 
considering only one nearby initial condition, joined by some small deviation
vector to the original point. In what follows, most of our simulations use
this faster (but less robust) \emph{deviation vector} method, but we have
checked many of the results using the Jacobian method. Further details of the
various techniques for calculating Lyapunov exponents appear
in~\cite{Hartl_2002_1} and especially~\cite{Hartl_thesis}. 

It is worth noting that we can think of the PN system as
\emph{constrained}, since we wish to think of the spin magnitudes as fixed. 
In other words, given an initial spin vector, a ``nearby'' initial spin
should point in a different direction but have the same magnitude. The
system thus has only ten true degrees of freedom (three for relative position
and momentum, and two for each spin), and should therefore have only ten
Lyapunov exponents. The constraints lead to significant complications in
calculating the Lyapunov exponents; see~\cite{Hartl_thesis} for several
methods of addressing these complications.

The principal value of the largest Lyapunov exponent is that it provides the
$e$-folding timescale $t_\lambda = 1/\lambda$ for the divergence of nearby
trajectories. The formal definition of $\lambda$ in Eq.~(\ref{eq:lyap})
requires an infinite-time limit, but of course any numerical method for
$\lambda$ must introduce some finite cutoff. As a result, in general it is
impossible to say with any certainty that a system is not chaotic---even if
it appears that $\lambda_\mathrm{max} \rightarrow 0$ for some
$t_\mathrm{cutoff}$, chaos may yet manifest itself on longer timescales. 
Nevertheless, it is possible to calculate nonchaotic \emph{baseline orbits}
(corresponding, for example, to the PN terms through the spin-orbit
coupling), whose nearby initial conditions still exhibit some nonzero
(power-law) separation. If a suspected chaotic orbit has a Lyapunov exponent
with a magnitude similar to a baseline orbit, we say that it is
indistinguishable from or consistent with zero, and hence is probably not
chaotic.

For the present problem, the most relevant timescale is the inspiral time due
to the energy loss from gravitational radiation. For the quasicircular
orbits considered below, this is approximately given by the following
formula~\cite{Peters1964},
\begin{equation}
t_\mathrm{insp} = \frac{5}{256}\,\frac{M^2}{\mu}\,\left(\frac{r}{M}\right)^4,
\end{equation}
where $M$~is the total mass and~$\mu$ is the reduced mass. For eccentric 
orbits we use Eq.~(5.14) of ~\cite{Peters1964}. 
We then adopt the
criterion $t_\lambda < t_\mathrm{insp}$ as an operational definition of
chaos, which is equivalent to the condition
\begin{equation}
\lambda \, t_\mathrm{insp} > 1\qquad\mbox{condition for chaotic orbit}.
\end{equation}
On the other hand, if $t_\lambda > t_\mathrm{insp}$, then, even if the system
is formally chaotic in the conservative limit we consider here, the chaos
will not have time to manifest itself before the final plunge.

\subsection{A survey of quasicircular orbits}
\label{sec:quasicircular_survey}

In this section, we elucidate the effects of varying the parameters in the PN
equations of motion on the presence of chaos in the resulting dynamics. In
many of the examples, we parametrize the orbits by their radii, or,
equivalently, by the ``gravitational wave frequency'':
\begin{equation}
\label{eq:fGW}
 f_\mathrm{GW}^{\rm Newt} = \frac{1}{\pi}\,\left(\frac{GM}{r^3}\right)^{1/2},
\end{equation}
where we restore the factor of~$G$ so that the result is in~Hz. It is
essential to note that Eq.~(\ref{eq:fGW}) is the \emph{Newtonian}
gravitational wave frequency, which is valid only for radii that satisfy
$r\gg M$. Nevertheless, Eq.~(\ref{eq:fGW}) provides a convenient way to
parameterize the initial conditions by radius in a way that has transparent
physical significance in the nonrelativistic limit. When we refer below to
an orbit with gravitational wave frequency of (say) 240~Hz, we mean an orbit
with a radius that satisfies Eq.~(\ref{eq:fGW}) when
$f_\mathrm{GW}^{\rm Newt}=240~\textrm{Hz}$. It is important to remember that this is
not in general the true frequency of the gravitational wave---for example,
at 2PN order, for equatorial orbits, averaging over an orbit yields~\cite{Kidder,BC,BI}

\begin{widetext}
\begin{eqnarray}
 f^{\rm 2PN}_\mathrm{GW} &=& \frac{1}{\pi}\,\left(\frac{GM}{r^3}\right)^{1/2} \left \{1 + 
\frac{1}{2}(-3 + \eta)\frac{GM}{r} - \frac{1}{2} \frac{1}{M^2}\widehat{\mathbf{L}} \cdot \mathbf{S}_{\rm eff} 
\left (\frac{GM}{r}\right )^{3/2} + \frac{1}{2} \left (3 + \frac{7}{8} \eta + \frac{3}{4}\eta^2 
\right )\left (\frac{GM}{r}\right )^2 + \right . \nonumber \\
&& \left . - \frac{3}{4} \frac{1}{\eta}\frac{1}{M^4} 
\left [(\mathbf{S}_1 \cdot \mathbf{S}_2) - 3 (\widehat{\mathbf{L}} \cdot 
\mathbf{S}_1) (\widehat{\mathbf{L}} \cdot \mathbf{S}_2) \right ]\left (\frac{GM}{r}\right )^2 \right \}\, 
\label{fPN}
\end{eqnarray}
\end{widetext}
with $\widehat{\mathbf{L}} = \mathbf{L}/|\mathbf{L}|$. For a binary with
total mass $(10+10)\,M_\odot$ at radius $r = 5.5658\,M$ 
($f_\mathrm{GW}^\mathrm{Newt} = 240$~Hz), Eq.~(\ref{fPN}) gives $f_{\rm
GW}^{\rm 2PN} = \{181, 194, 212\}$~Hz when spins are aligned with (orbital)
angular momentum, zero, and anti-aligned with angular momentum, respectively.
For the same spin orientations at $r = 5.742\,M$
($f_\mathrm{GW}^\mathrm{Newt} = 150$~Hz), Eq.~(\ref{fPN}) gives $f_{\rm
GW}^{\rm 2PN} = \{122, 127, 134\}$~Hz. 

\subsubsection{A chaotic quasicircular orbit}

We find that post-Newtonian orbits satisfying the conditions for
quasicircularity (Sec.~\ref{sec:quasicircular}) can be chaotic (though
\emph{only} for  rather small initial radial separations), as shown in
Figs.~\ref{fig:PN_orbit}--\ref{fig:lyapunov_nonzero}. Note from
Fig.~\ref{fig:PN_Lyapunov} that the Lyapunov exponents come in $\pm\lambda$
pairs, a characteristic of Hamiltonian dynamical systems. Note also that the
principal exponent calculated using the deviation vector method
(Fig.~\ref{fig:lyapunov_nonzero}) agrees closely with the largest exponent
determined from Fig.~\ref{fig:PN_Lyapunov}, which uses the more complicated
Jacobian method to find the exponents.

\subsubsection{Varying spin directions}
\label{sec:varying_directions}

\begin{figure}
\includegraphics[width=3.in]{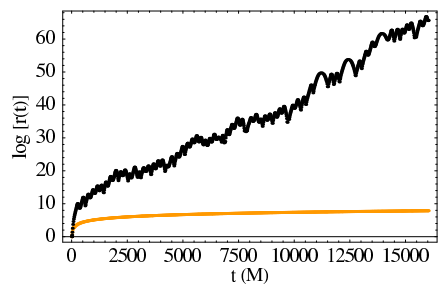}
\caption{The natural logarithm of the principal ellipsoid axis vs.\ time for
the system shown in Fig.~\ref{fig:PN_orbit}, using the deviation vector 
method. The slope of the line is the Lyapunov exponent, which is
approximately $\lambda = 3.2\times10^{-3}\,M^{-1}$ using a least-squares fit
(which agrees closely with the value from Fig.~\ref{fig:PN_Lyapunov}, which
uses the more sophisticated Jacobian method to find the exponent). This
corresponds to a Lyapunov ($e$-folding) timescale of $t_\lambda = 1/\lambda =
3.1\times10^2\,M$, which is less than a fifth of the inspiral timescale. The
simulation data for a nonchaotic orbit (light) is shown for reference.}
\label{fig:lyapunov_nonzero}
\end{figure}

\begin{figure}
\includegraphics[width=3.in]{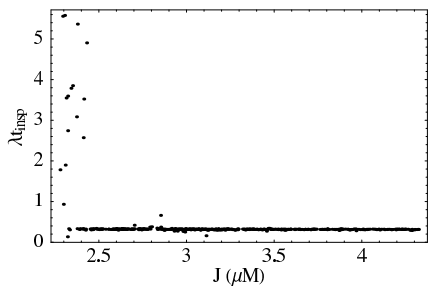}
\caption{Lyapunov exponents for 500 quasicircular orbits as a function of
total angular momentum~$J$ for the $(10+10) M_\odot$ configuration. The spin
for each body is maximal with random initial spin angles $(\theta, \phi)$
(which overweights the poles). The initial radius
corresponds to a gravitational wave frequency of $f_\mathrm{GW}^{\rm Newt} =
240\,\textrm{Hz}$. The Lyapunov exponents are measured in terms of the
inverse inspiral time $1/t_\mathrm{insp}$, so that $\lambda t_\mathrm{insp}
> 1$ indicates that nearby trajectories diverge by a factor of~$e$ on a
timescale shorter than the inspiral timescale. There are 14 such chaotic
initial conditions out of the 500 orbits considered for this configuration. 
They are clustered at the low end of the angular momentum range, indicating
that the spins are closely aligned with each other and anti-aligned with the
orbital angular momentum.}
\label{fig:Lyapunov_J_10_10}
\end{figure}

\begin{figure}
\includegraphics[width=3.in]{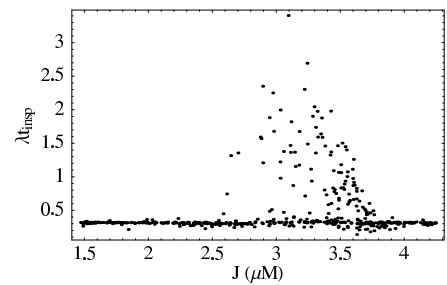}
\caption{Lyapunov exponents for 500 quasicircular orbits as a function of
total angular momentum~$J$ for the $(20+10) M_\odot$ configuration. The spin
for each body is maximal and randomly oriented, and the initial radius
corresponds to a gravitational wave frequency of $f_\mathrm{GW}^{\rm Newt} =
240\,\textrm{Hz}$. The Lyapunov exponents are measured in terms of the
inverse inspiral time $1/t_\mathrm{insp}$, so that $\lambda t_\mathrm{insp}
> 1$ indicates that nearby trajectories diverge by a factor of~$e$ on a
timescale shorter than the inspiral timescale. There are 49 such chaotic
initial conditions out of the 500 orbits considered for this configuration. 
Unlike the $(10+10) M_\odot$ case shown in Fig.~\ref{fig:Lyapunov_J_10_10},
the chaotic orbits in this case correspond to total angular momentum in the
middle of the range.}
\label{fig:Lyapunov_J_20_10}
\end{figure}

We illustrate the effect of varying the spin directions by generating a large
number of quasicircular orbits with randomly oriented (maximal) spins. For
each spin configuration, we choose the radius corresponding to a
gravitational wave with frequency $f_\mathrm{GW}^{\rm Newt} =
240\,\textrm{Hz}$ (the high end of the LIGO/VIRGO frequency band). The choice
of radius is motivated by two main factors. First, choosing the lowest
possible radius (consistent with the abilities to detect the corresponding
gravitational waves) likely represents a worst-case scenario for chaos, since
low-radius regions correspond to stronger nonlinearities in the equations of
motion (as noted in~\cite{Hartl_2002_2}). Second, minimizing the radius
minimizes the inspiral timescale~$t_\mathrm{insp}$, which in turn minimizes
the computational cost of a final integration time significantly longer than
$t_\mathrm{insp}$. This allows us to achieve a better bound on the suspected
zero Lyapunov exponent, and increases our confidence that apparent nonzero
exponents represent genuine chaotic behavior. In what follows, the final
integration time is ten times the inspiral time of each orbit.

\begin{table*}
\caption[The prevalence of chaos in post-Newtonian quasicircular orbits at
240~Hz]
{\label{table:random_spins_240Hz}
The prevalence of chaos in post-Newtonian quasicircular orbits at 240~Hz,
for spin directions chosen randomly on a unit sphere. We calculate the
fraction of orbits whose $e$-folding times $t_\lambda = 1/\lambda$ are less
than the inspiral time~$t_\mathrm{insp}$, which is our operational definition
of chaos. The final integration time is ten times the inspiral time. We also
include 95\% confidence intervals for the reported fractions, and we show the
average value of $\lambda$ measured in units of the inverse inspiral time for
the $(20+10) M_\odot$ configuration (the only case in our simulation with
more than one chaotic orbit). The simulation data represent 500 randomly
chosen initial spin directions for each configuration, with the initial
radius fixed by requiring a gravitational wave frequency of 240~Hz (as
determined by the Newtonian formula).
}
\medskip
\begin{center}
\begin{tabular}{|c|c|c|c|}\hline
Configuration & Fraction chaotic & 95\% confidence interval 
 & Average chaotic $\lambda t_\mathrm{insp}$
 \\ \hline
$(20+5) M_\odot$ & 0 & $[0,\,0.00738]$ & \\
$(10+5) M_\odot$ & 0 & $[0,\,0.00738]$ & \\
$(5+5) M_\odot$ & 0 & $[0,\,0.00738]$ & \\
$(10+10) M_\odot$ & $0.002$ & $[5.06\times10^{-5},\,0.0111]$ & \\
$(20+20) M_\odot$ & 0 & $[0,\,0.00738]$ & \\
$(20+10) M_\odot$ & $0.104$ & $[0.0777,\,0.136]$ & 1.45 \\ 
$(15+5) M_\odot$ & 0 & $[0,\,0.00738]$ & \\ 
\hline
\end{tabular}
\end{center}
\end{table*}

We consider the following mass configurations: $(20+5) M_\odot$, $(10+5)
M_\odot$, $(5+5) M_\odot$, $(10+10) M_\odot$, $(20+20) M_\odot$, $(20+10)
M_\odot$, and $(15+5) M_\odot$. The result of choosing $N = 500$ randomly
oriented initial spins for each case appears in
Table~\ref{table:random_spins_240Hz}. We find the presence of chaotic orbits
for the $(10+10) M_\odot$ and $(20+10) M_\odot$ cases, but we find no chaos
for any other configuration. Fig.~\ref{fig:lyapunov_nonzero} shows a Lyapunov
plot for the strongest chaos in our simulation data. The onset of chaos is
marked by a transition from linear (or at most power-law) separation of
nearby initial conditions to exponential separation. On our Lyapunov plot
(which is logarithmic on its vertical axis), chaotic orbits appear as linear
growth.

Our initial simulation for the $(10+10) M_\odot$ configuration used random
spin angles $(\theta, \phi)$, which does not correspond to a random
orientation but rather overweights the poles. This was a stroke of good
luck: as shown in Fig.~\ref{fig:Lyapunov_J_10_10}, for $(10+10) M_\odot$ the
chaotic orbits are clustered at the lowest values of the total angular
momentum~$J$, corresponding to initial spin vectors nearly anti-aligned with
the orbital angular momentum~$\mathbf{L}$ (so that $J = |\mathbf{L} +
\mathbf{S}|$ is minimized). When running the simulation again using
randomly oriented spins, we find only one chaotic orbit for this
configuration. The association of chaos with low values of $\mathbf{J}$ holds
also for the sole chaotic orbit found in the $(10+5) M_\odot$ case, but it is
not a general result: as Fig.~\ref{fig:Lyapunov_J_20_10} shows, chaos for the
$(20+10) M_\odot$ configuration occurs mainly for values of~$J$ in the middle
of the possible range.

\subsubsection{Varying initial frequencies}
\label{sec:varying_frequencies}

\begin{figure}
\includegraphics[width=3.in]{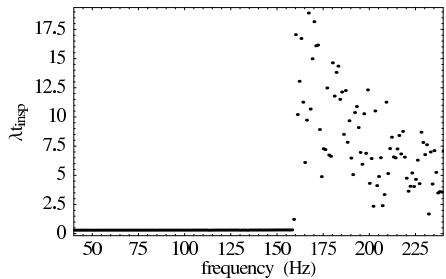}
\caption{Lyapunov exponents as a function of the (Newtonian)
gravitational-wave frequency for quasicircular orbits of two maximally
spinning $10\,M_\odot$ black holes.  The Lyapunov exponents are measured in
terms of the inverse inspiral time $1/t_\mathrm{insp}$, and the frequencies
are chosen to correspond closely to the LIGO/VIRGO frequency band. There is
an abrupt transition to chaos at approximately 160~Hz.}
\label{fig:lyapunov_frequency}
\end{figure}

\begin{figure}
\includegraphics[width=3.in]{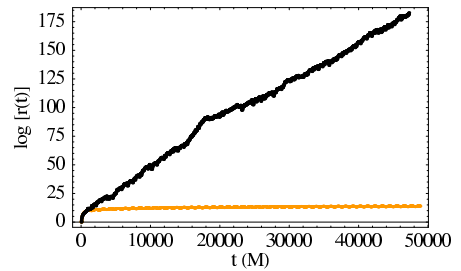}
\caption{The natural logarithms of the principal ellipsoid axes vs.\ time 
for frequencies on opposite sides of the transition to chaos shown in
Fig.~\ref{fig:lyapunov_frequency}: $f_\mathrm{GW}^{\rm Newt} = 158~\textrm{Hz}$ (light)
and $f_\mathrm{GW}^{\rm Newt} = 160~\textrm{Hz}$ (dark). The slope of each line is the
Lyapunov exponent, with $\lambda \approx 7\times10^{-5}\,M^{-1}$ (light,
nonchaotic/consistent with zero) and $\lambda \approx
4\times10^{-3}\,M^{-1}$ (dark, chaotic). The orbits corresponding to these
two frequencies appear in Figs.~\ref{fig:lyapunov_frequency_nonchaotic}
and~\ref{fig:lyapunov_frequency_chaotic}.}
\label{fig:lyapunov_frequency_compare}
\end{figure}

We now investigate the results of varying the initial (gravitational-wave)
frequencies for the strongest chaotic orbit from the previous section (as
illustrated in Fig.~\ref{fig:lyapunov_nonzero}). The result appears in
Fig.~\ref{fig:lyapunov_frequency}, which shows that, even for this worst-case
scenario, i.e., the system with the largest Lyapunov exponent, chaos is
absent for initial radii corresponding to $f_\mathrm{GW}^{\rm Newt}$ less
than 160~Hz.  Above 160~Hz, there is an abrupt change in the dynamics from
regular to chaotic (Fig.~\ref{fig:lyapunov_frequency_compare}), with a
maximum Lyapunov exponent more than 18 times the inverse inspiral time
(meaning that nearby trajectories diverge by a factor of $e$~in a time
$t_\lambda \approx t_\mathrm{insp}/18$).

Cornish and Levin have recently pointed out~\cite{CL2003} that if
unstable orbits are perturbed they could become locus of chaos. The abrupt
transition from nonchaotic to chaotic behavior that we observe in
Fig.~\ref{fig:lyapunov_frequency}, when we increase the GW frequency (and
hence lower the radial separation), could correspond to the transition from
stable to unstable orbits in the spinning ADM Hamiltonian. Thus, to better
understand the onset of chaos, it would be worthwhile to explore this
intuition further by applying some stability criterion in the Hamiltonian
framework of the kind worked out in the non-spinning case by~\cite{BI}.

\begin{figure}
\includegraphics[width=3.in]{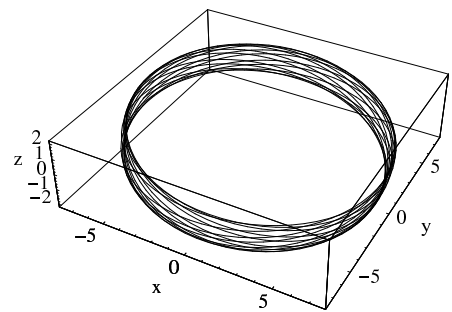}
\caption{The nonchaotic quasicircular orbit of two maximally spinning
$10\,M_\odot$ black holes, corresponding
to a gravitational wave frequency of $f_\mathrm{GW}^{\rm Newt} = 158~\textrm{Hz}$. 
The spins are the same
as in Fig.~\ref{fig:PN_orbit}.
The
corresponding Lyapunov plot is shown in 
Fig.~\ref{fig:lyapunov_frequency_compare}. The orbit's radius is
approximately constant, as required for a true quasicircular orbit.}
\label{fig:lyapunov_frequency_nonchaotic}
\end{figure}

\begin{figure}
\includegraphics[width=3.in]{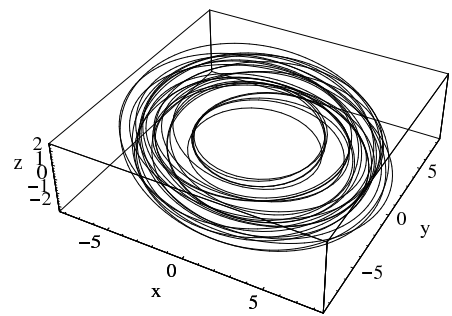} \caption{The
chaotic quasicircular orbit of two maximally spinning $10\,M_\odot$ black
holes, corresponding to a gravitational
wave frequency of $f_\mathrm{GW}^{\rm Newt} = 160~\textrm{Hz}$. The spins are the same
as in Fig.~\ref{fig:PN_orbit}.
The corresponding
Lyapunov plot is shown in Fig.~\ref{fig:lyapunov_frequency_compare}. Note
that the quasicircularity has broken down; the radius is not even
approximately constant. This qualitative change in the orbit accompanies the
onset of chaos as the frequency increases (with a corresponding decrease in
radius), as illustrated in Fig.~\ref{fig:lyapunov_frequency}.}
\label{fig:lyapunov_frequency_chaotic}
\end{figure}

In this particular case, the qualitative change in the dynamical behavior
from nonchaotic to chaotic is mirrored in the orbits themselves. In
particular, the onset of chaos is associated with a breakdown in the
quasicircularity of the orbit. As shown in
Fig.~\ref{fig:lyapunov_frequency_nonchaotic}, just below the transition to
chaos the orbit is nearly circular. Just above the transition, despite
satisfying the conditions for quasicircularity at initial time, the orbits are not even
approximately circular along the evolution 
because of the strong spin coupling, as shown in
Fig.~\ref{fig:lyapunov_frequency_chaotic}

\begin{table}
\caption[The prevalence of chaos in post-Newtonian quasicircular orbits at
40~Hz]
{\label{table:random_spins_40Hz}
The prevalence of chaos in post-Newtonian quasicircular orbits at 40~Hz, for
spin directions chosen randomly on a unit sphere. We calculate the fraction
of orbits whose $e$-folding times $t_\lambda = 1/\lambda$ are less than the
inspiral time~$t_\mathrm{insp}$, which is our operational definition of
chaos. The final integration time is ten times the inspiral time. We also
include 95\% confidence intervals for the reported fractions. The simulation
data represent 500 randomly chosen initial spin directions for each
configuration, with the initial radius fixed by requiring a gravitational
wave frequency of 40~Hz (as determined by the Newtonian formula). In this
case, the number of chaotic orbits for each configuration is zero.
}
\medskip
\begin{center}
\begin{tabular}{|c|c|c|}\hline
Configuration & Fraction chaotic & 95\% confidence interval \\ \hline
$(20+5) M_\odot$ & 0 & $[0,\,0.00738]$ \\
$(10+5) M_\odot$ & 0 & $[0,\,0.00738]$ \\
$(5+5) M_\odot$ & 0 & $[0,\,0.00738]$ \\
$(10+10) M_\odot$ & 0 & $[0,\,0.00738]$ \\
$(20+20) M_\odot$ & 0 & $[0,\,0.00738]$ \\
$(20+10) M_\odot$ & 0 & $[0,\,0.00738]$ \\ 
$(15+5) M_\odot$ & 0 & $[0,\,0.00738]$ \\ 
\hline
\end{tabular}
\end{center}
\end{table}

To verify that the disappearance of chaos at lower frequencies is generic, we
repeated the 240~Hz survey for the lower end of the frequency range
considered ($f_{GW}^{\rm Newt} = 40~\textrm{Hz}$). The inspiral times are very long in
this case, requiring patience on the part of the simulator, but the results
are gratifying: as shown in Table~\ref{table:random_spins_40Hz}, we found not
even one orbit with a Lyapunov time less than the inspiral time at 40~Hz. 
Any chaos, if present, manifests itself in this case on timescales longer
than $t_\mathrm{insp}$.

\subsubsection{Varying spin magnitudes}
\label{sec:varying_spin}

In Sec.~\ref{sec:varying_frequencies}, we created a one-parameter family of
orbits by taking the worst offender from Sec.~\ref{sec:varying_directions}
and varying the frequency (or, equivalently, the quasicircular radius). In
this section, we do the same, but fix the frequency and vary the magnitude of
one of the spins. 

\begin{figure}
\includegraphics[width=3.in]{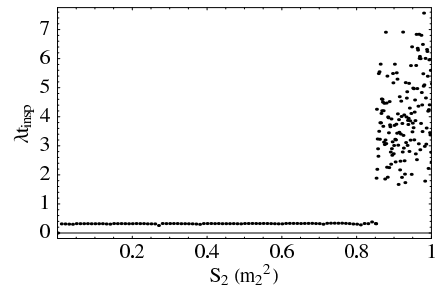}
\caption{Lyapunov exponents as a function of spin for quasicircular orbits
of two $10\,M_\odot$ black holes. We fix the spin of one hole at the maximum
value ($S_1 = m_1^2$), and also fix the spin directions (which are the same as
in Fig.~\ref{fig:PN_orbit}), and then vary the
spin~$S_2$ of the second body. The Lyapunov exponents are measured in terms
of the inverse inspiral time $1/t_\mathrm{insp}$. There is an abrupt
transition to chaos when~$S_2$ exceeds~$0.85$ (measured in units of
$m_2^2$).}
\label{fig:lyapunov_spin}
\end{figure}

We showed in Sec.~\ref{sec:varying_directions} that the worst-offender orbit
(with $m_1=m_2 = 10\,M_\odot$) is chaotic when $S_2 = 1$ (measured in units
of~$m_2^2$). As shown in Fig.~\ref{fig:lyapunov_spin}, the dynamics are
nonchaotic for most values of $S_2$, with a transition to chaos at
approximately $S=0.85$. Since the dynamics are nonchaotic when $S_2=0$, the
chaos must be produced by the spin terms in the Hamiltonian.

\begin{figure}
\includegraphics[width=3.in]{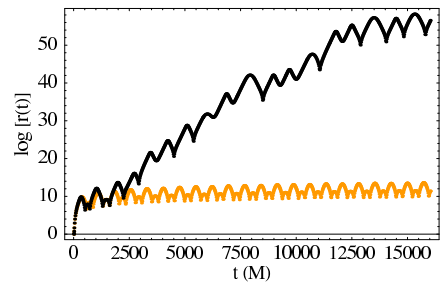}
\caption{The natural logarithms of the principal ellipsoid axes vs.\ time 
for values of $S_2$ on opposite sides of the transition to chaos shown in
Fig.~\ref{fig:lyapunov_spin}: $S_2 = 0.83$ (light) and $S_2 = 0.86$ (dark).
The slope of each line is the Lyapunov exponent, with $\lambda \approx
2.0\times10^{-4}\,M^{-1}$ (light, nonchaotic/consistent with zero) and
$\lambda \approx 3.5\times10^{-3}\,M^{-1}$ (dark, chaotic). The orbits
corresponding to these two spins appear in
Figs.~\ref{fig:lyapunov_spin_nonchaotic}
and~\ref{fig:lyapunov_spin_chaotic}.}
\label{fig:lyapunov_spin_compare}
\end{figure}

\begin{figure*}
\begin{tabular}{ccc}
\includegraphics[width=3.in]{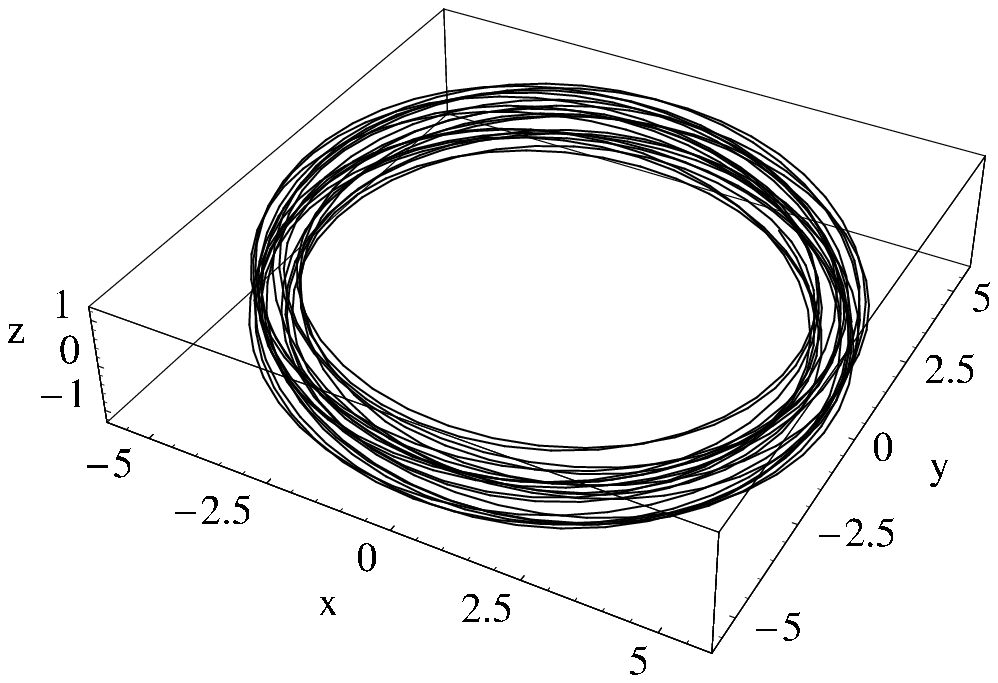} & \hspace{0.5in}
	& \includegraphics[width=3.in]{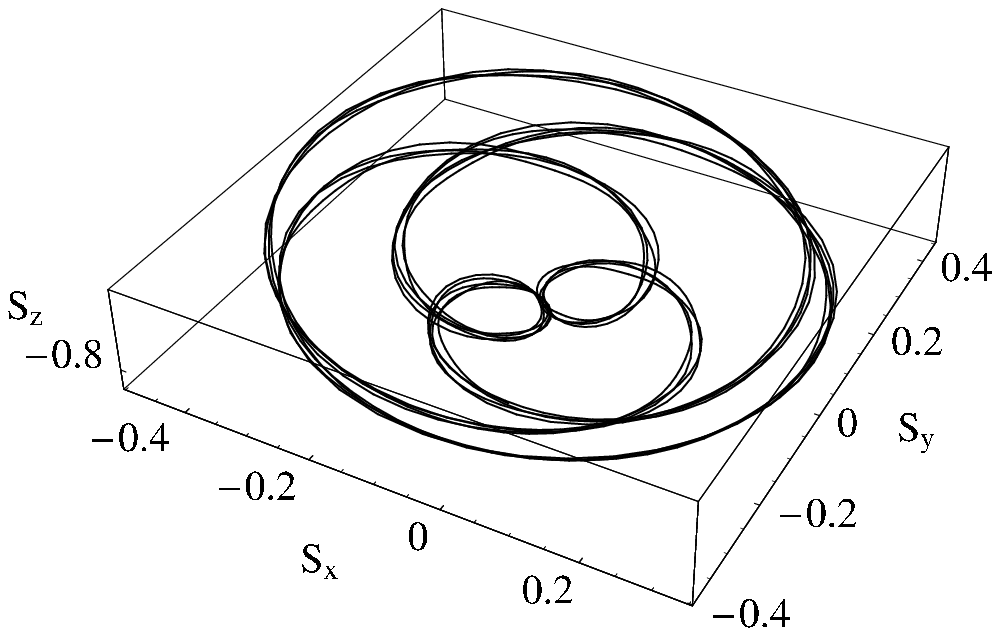}\\
(a) & & (b)\medskip\\
\end{tabular}
\caption{(a) The nonchaotic quasicircular orbit of two $10\,M_\odot$ black
holes, with spins magnitudes of $S_1 =
1$ and $S_2 = 0.83$; (b) Cartesian ``spin space'' showing the time-evolution
of $S_x$, $S_y$, and $S_z$. Compare with
Fig.~\ref{fig:lyapunov_spin_chaotic}(b). The corresponding Lyapunov plot is
shown in Fig.~\ref{fig:lyapunov_spin_compare}.}
\label{fig:lyapunov_spin_nonchaotic}
\end{figure*}

\begin{figure*}
\begin{tabular}{ccc}
\includegraphics[width=3.in]{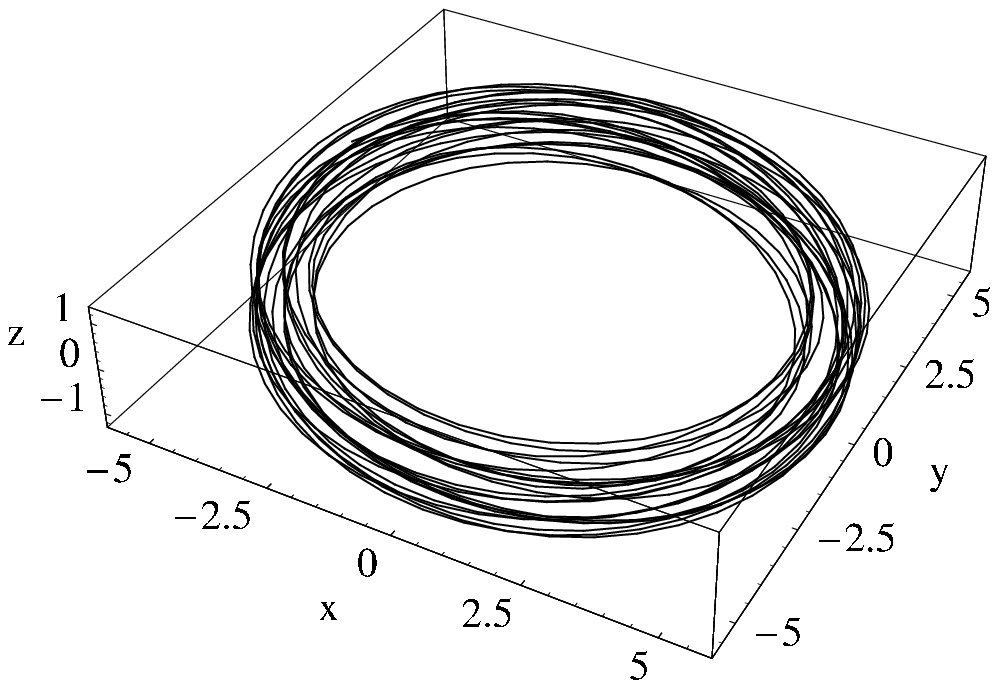} & \hspace{0.5in}
	& \includegraphics[width=3.in]{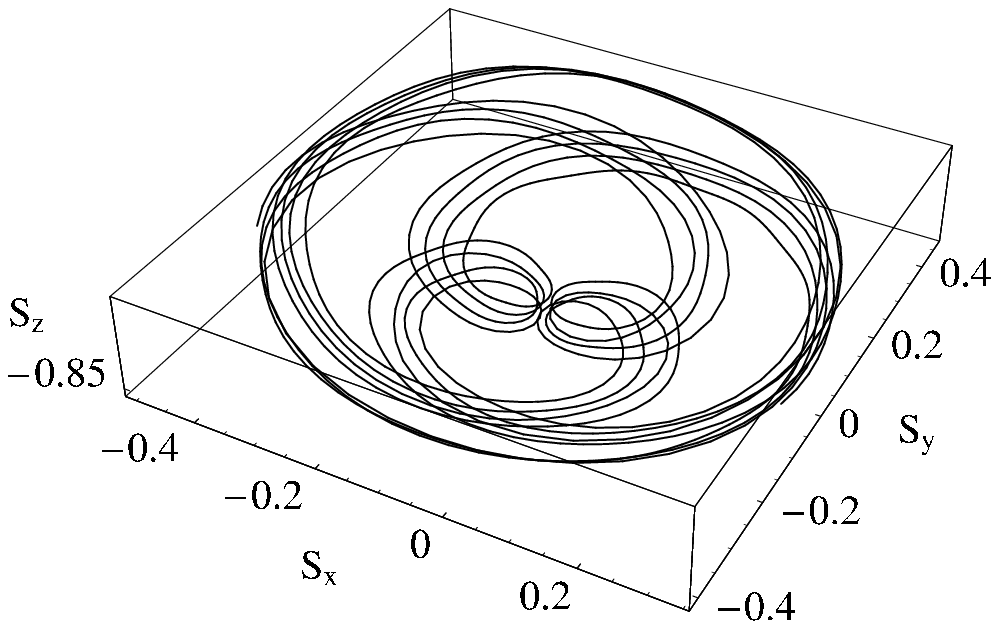}\\
(a) & & (b)\medskip\\
\end{tabular}
\caption{(a) The chaotic quasicircular orbit of two $10\,M_\odot$ black
holes, with spins magnitudes of $S_1 =
1$ and $S_2 = 0.86$; (b) Cartesian ``spin space'' showing the time-evolution
of $S_x$, $S_y$, and $S_z$. The corresponding Lyapunov plot is shown in 
Fig.~\ref{fig:lyapunov_spin_compare}. There appears to be no qualitative
difference between the orbit~(a) and the orbit in
Fig.~\ref{fig:lyapunov_spin_nonchaotic}(a), but there is a qualitative change
in the spin behavior. Unlike the frequency transition shown in
Fig.~\ref{fig:lyapunov_frequency_compare}, which gives a qualitative change
in the evolution of the spatial variables, the spin transition to chaos
manifests primarily itself in the spin degrees of freedom.}
\label{fig:lyapunov_spin_chaotic}
\end{figure*}

In Fig.~\ref{fig:lyapunov_spin_compare}, we show Lyapunov plots for orbits on
either side of the chaotic transition. Although the difference is not as
dramatic as the frequency-induced transition
(Fig.~\ref{fig:lyapunov_frequency_compare}), there is still a qualitative
change in the value of the principal Lyapunov exponent. Unlike the system in
Sec.~\ref{sec:varying_directions}, this transition does not give rise to a
qualitative change in the orbit as the spin is varied. Instead, the chaos
manifests itself in the time-evolution of the spins, as shown in
Figs.~\ref{fig:lyapunov_spin_nonchaotic} and~\ref{fig:lyapunov_spin_chaotic}.

\subsubsection{Varying the PN terms}

The previous results in this section included all the PN terms described in
Sec.~\ref{sec:Hamiltonian}, with the exception of the 3PN term. Here we
investigate the effect of varying these terms. We first consider the effect
of the 3PN term, and then investigate the effects of turning off one or more
of the spin terms. 

\begin{figure}
\includegraphics[width=3.in]{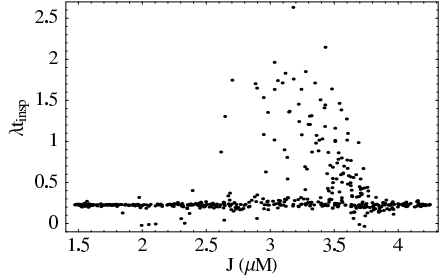}
\caption{Lyapunov exponents for 500 quasicircular orbits as a function of
total angular momentum~$J$ for the $(20+10) M_\odot$ configuration with 3PN
term added. The initial conditions are identical to those in
Fig.~\ref{fig:Lyapunov_J_20_10}; the prevalence of chaos is not strongly
affected by the presence of the 3PN term.}
\label{fig:Lyapunov_J_20_10_3PN}
\end{figure}

\begin{figure}
\includegraphics[width=3.in]{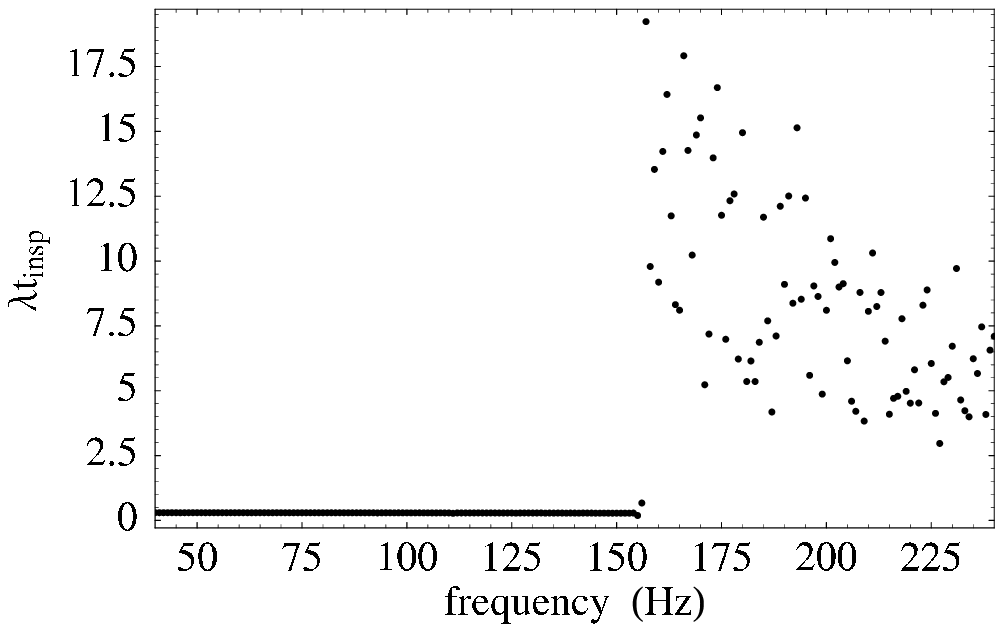}
\caption{Lyapunov exponents as a function of (Newtonian) gravitational-wave frequency for
quasicircular orbits of two maximally spinning $10\,M_\odot$ black holes with
3PN term added. 
The Lyapunov exponents are measured in terms of the inverse inspiral time
$1/t_\mathrm{insp}$, and the frequencies are chosen to correspond closely to
the LIGO frequency band. There is an abrupt transition to chaos at
approximately 155~Hz. The initial conditions are identical to those in
Fig.~\ref{fig:lyapunov_frequency}.}
\label{fig:lyapunov_frequency_3PN}
\end{figure}

\begin{figure*}
\begin{tabular}{ccc}
\includegraphics[width=3.in]{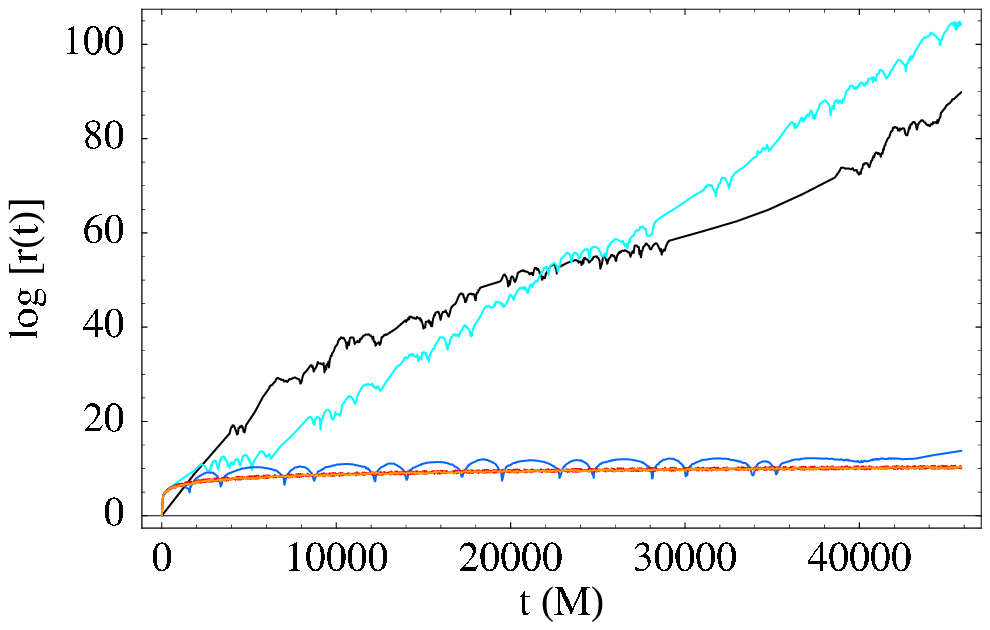} & \hspace{0.5in}
	& \includegraphics[width=3.in]{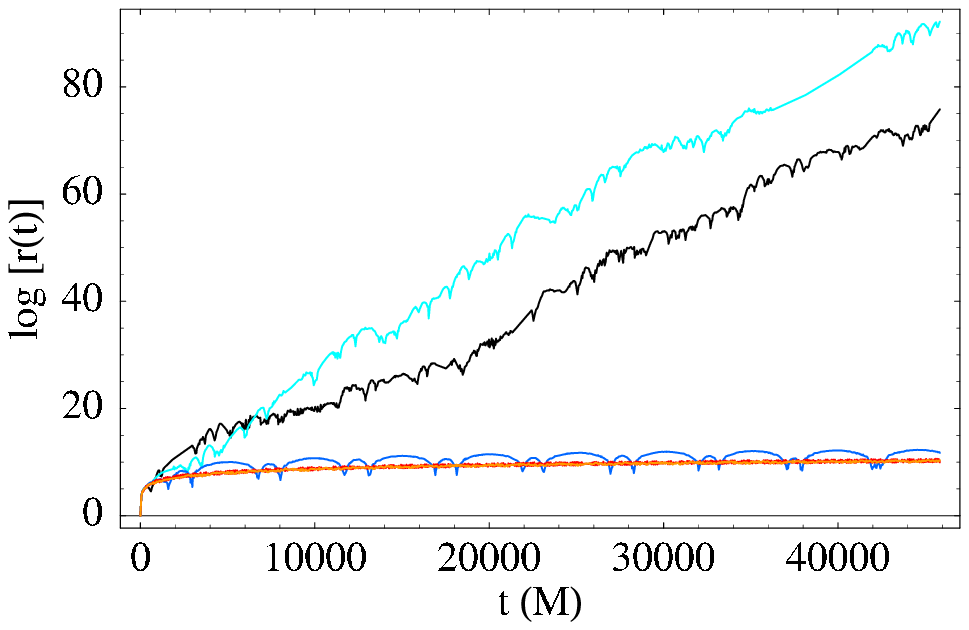}\\
(a) & & (b)\medskip\\
\end{tabular}
\caption{
\label{fig:lyapunov_PNTerms}
The natural logarithms of the longest ellipsoid axis~$r$ vs.~$t$
for the strongest chaos found in the $(20+10) M_\odot$ configuration (as
illustrated in Fig.~\ref{fig:Lyapunov_J_20_10}), for varying post-Newtonian
terms: (a) 3PN turned off; (b) 3PN turned on. The final time is 50~times
the inspiral time, and the slopes of the lines are the Lyapunov exponents. 
In all cases, the Newtonian, 1PN, 2PN, and spin-orbit terms are turned on,
but some of the others may be turned off; from top to bottom (with colors,
visible in electronic versions of this paper, noted parenthetically): all PN
terms (black); $S_1S_1$ and $S_2S_2$ turned off (cyan); $S_1S_1$ turned off
(blue, nonchaotic); $S_2S_2$ turned off (red, nonchaotic); $S_1S_1$,
$S_2S_2$, and $S_1S_2$ turned off (orange, nonchaotic). For this particular
case, the $S_1S_2$ spin-coupling term is necessary for chaos to appear.
}
\end{figure*}

\begin{figure}
\begin{center}
\includegraphics[width=3in]{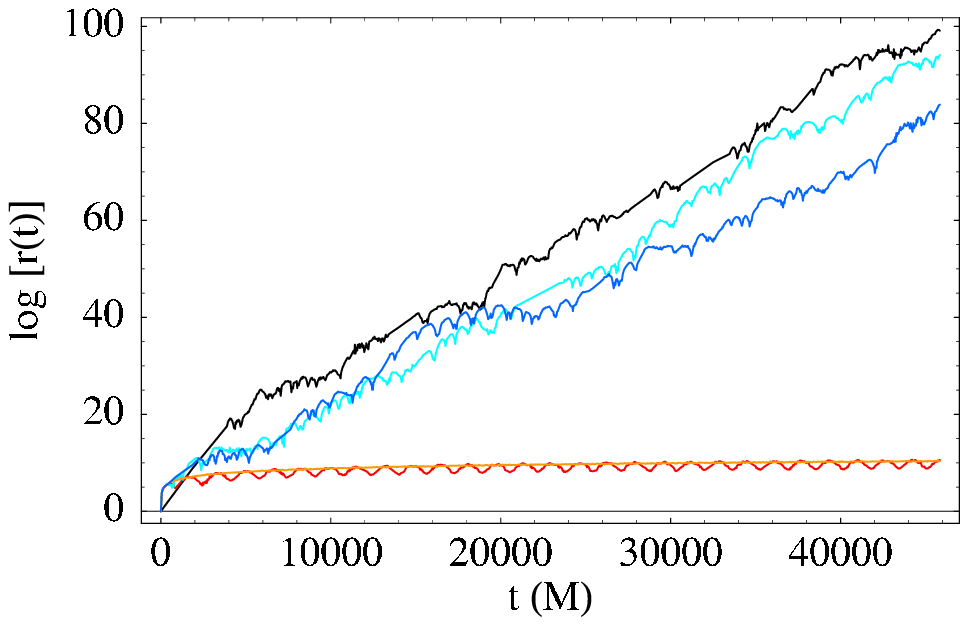}
\end{center}
\caption
{\label{fig:lyapunov_PNTerms2}
The natural logarithms of the longest ellipsoid axis~$r$ vs.~$t$ for a
strongly chaotic $(20+10) M_\odot$ configuration, for varying post-Newtonian
terms. The initial spins (both maximal) are $\mathbf{S}_1= (-0.935125,
0.329567, 0.130101)\,m_1^2$ and $\mathbf{S}_2= (0.039523, -0.54303,
-0.838783)\,m_2^2$, while the initial radius is $r=4.318\,M$, corresponding to
$f_\mathrm{GW}^{\rm Newt} = 240~\textrm{Hz}$; the other initial conditions are fixed by
the conditions for quasicircularity (Sec.~\ref{sec:quasicircular}). The final
time is 50~times the inspiral time, and the slopes of the lines are the
Lyapunov exponents. In all cases, the Newtonian, 1PN, 2PN, and spin-orbit
terms are turned on, but some of the others may be turned off; from top to
bottom (with colors, visible in electronic versions of this chapter, noted
parenthetically): all PN terms (black); $S_1S_1$ and $S_2S_2$ turned off
(cyan); $S_1S_1$ turned off (blue); $S_2S_2$ turned off (wiggly/red,
nonchaotic); $S_1S_1$, $S_2S_2$, and $S_1S_2$ turned off (straight/orange,
nonchaotic).}
\end{figure}

In order to evaluate the effect of the 3PN term on chaos in the
post-Newtonian equations, we re-calculate the simulation of 500 orbits in the
$(20+10)M_\odot$ system shown in Fig.~\ref{fig:Lyapunov_J_20_10}; the result
appears in Fig.~\ref{fig:Lyapunov_J_20_10_3PN}. There are some minor
differences, including a slight suppression of the most chaotic orbit, but
the effect of 3PN is not strong. This result holds also for the $(10+10)
M_\odot$; in this case we examine the transition to chaos shown in
Fig.~\ref{fig:lyapunov_frequency}. As seen in
Fig.~\ref{fig:lyapunov_frequency_3PN}, the transition occurs at a slightly
lower frequency, but the difference between the two cases is small. Our
general conclusion is that the third post-Newtonian term has only minor
effects on the presence of chaos.

In order to investigate the effect of the spin terms in the post-Newtonian
Hamiltonian, we focus first on the strongest chaos found in our simulations
of the $(20+10) M_\odot$ configuration for $f_\mathrm{GW}^{\rm Newt} = 40\,\textrm{Hz}$
with 3PN turned on, as shown in Fig.~\ref{fig:Lyapunov_J_20_10}. The
Lyapunov plots for a variety of PN term combinations appears in
Fig.~\ref{fig:lyapunov_PNTerms}. The most important result is that the
spin-spin terms are crucial to the presence of chaos; when only Newtonian,
1PN, 2PN, 3PN and spin-orbit are turned on, the system is not chaotic. In
the case illustrated in Fig.~\ref{fig:lyapunov_PNTerms}, the $S_1S_2$ term by
itself causes chaos: the presence of either the spin quadrupole term is
irrelevant. This is not a general result: Fig.~\ref{fig:lyapunov_PNTerms2} 
shows a case where the $S_1S_1$ quadrupole term apparently exerts a
stabilizing influence: when only $S_2S_2$ is turned off, the system is
nonchaotic, but if $S_1S_1$ is then turned off as well the system returns to
chaotic behavior.

\subsection{Eccentric orbits}
\label{sec:eccentric_survey}

Although quasicircular orbits represent the most likely source of
gravitational waves from compact binaries detectable by ground-based
observatories, some sources may consist of binaries with non negligible 
eccentricity (such as produced, for example, by the Kozai mechanism~\cite{MH,LW}). 
It is therefore potentially relevant to investigate chaos for
eccentric orbits.

\subsubsection{A chaotic eccentric orbit}

In agreement with \cite{CornishLevin2002,CornishLevin2003}, we find that the
post-Newtonian equations can produce chaotic eccentric orbits; one example
appears in Fig.~\ref{fig:eccentric_lyapunov_plot}. It must be emphasized
that, as in the case of quasicircular orbits, the pericenter of the orbit
corresponding to Fig.~\ref{fig:eccentric_lyapunov_plot} is at $r_p = 5\,M$; 
thus the two black holes are rather close to each other and 
higher order PN corrections, especially higher spin 
PN corrections, should also be taken into account. 
Nevertheless, it is clear that the equations themselves can
produce chaotic solutions, even if those solutions have dubious physical
relevance.

\subsubsection{Parameter variation}

\begin{figure}
\begin{center}
\includegraphics[width=3in]{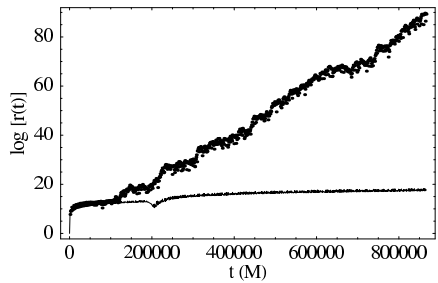}
\end{center}
\caption{\label{fig:eccentric_lyapunov_plot} The natural logarithms of the
longest ellipsoid axis~$r$ vs.~$t$ for a chaotic eccentric orbit. The slope
of the line gives $\lambda t_\mathrm{insp} = 1.01$, so that nearby
trajectories diverge by a factor of $e$ in approximately one inspiral time. 
The initial conditions
are $M_1 = 6$, $M_2 = 4$, $\mathbf{X} = (5, 0, 0)$, $\mathbf{P} = (0,
0.61644, 0.36160)$, $\mathbf{S}_1 = (-1.21570, -0.31859, 0.81886)$, and
$\mathbf{S}_2 = (-0.15273, 0.64525, -0.06902)$. For reference,
we show the corresponding plot for a
nonchaotic orbit (with the same initial conditions as above
except for $\mathbf{S}_1 = 0$).}
\end{figure}

\begin{figure*}
\begin{tabular}{ccc}
\includegraphics[width=3.in]{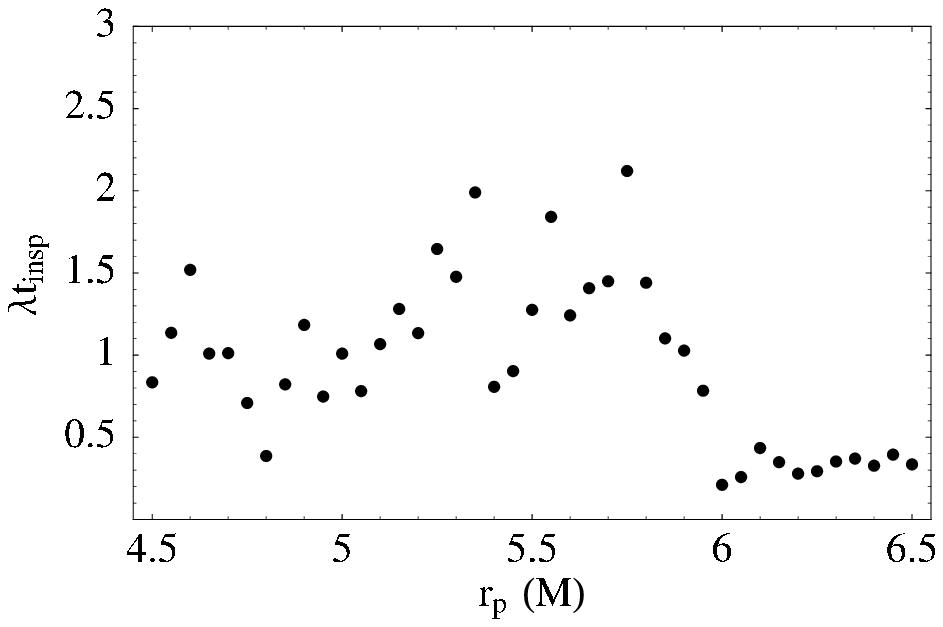} & \hspace{0.5in}
	& \includegraphics[width=3.in]{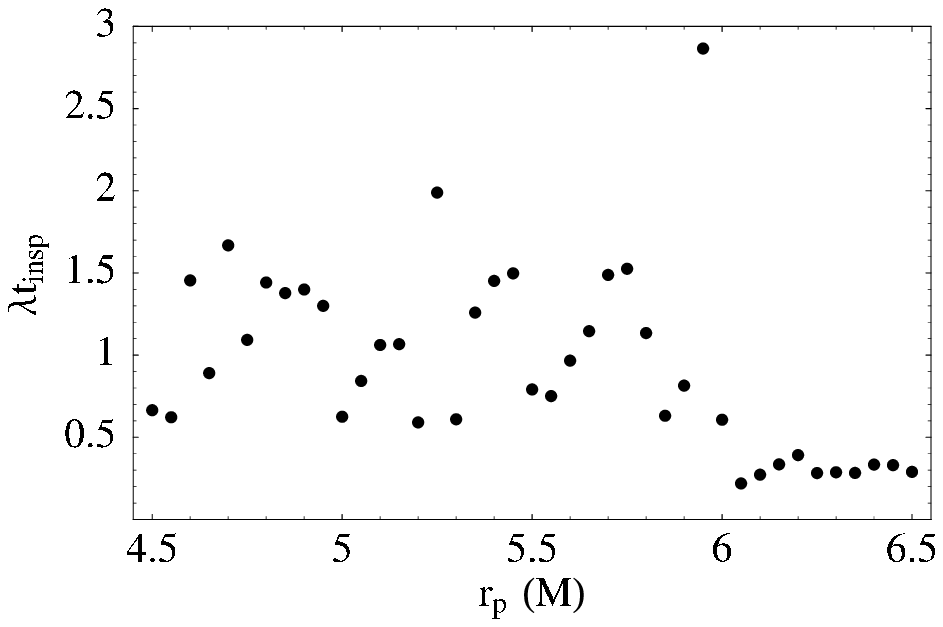}\\
(a) & & (b)\medskip\\
\end{tabular}
\caption{The dimensionless Lyapunov exponent $\lambda t_\mathrm{insp}$ as
a function of pericenter for (a) all terms through 2PN and (b) all terms
through 3PN. The initial spins are the same as those in
Fig.~\ref{fig:eccentric_lyapunov_plot}, with the other initial conditions
fixed by the parameterization method described in
Sec.~\ref{sec:eccentric}.
Note that the Lyapunov exponent decreases with increasing pericenter as
the spin coupling terms get smaller (consistent with the
the results in \cite{Hartl_2002_1,Hartl_2002_2}).
}
\label{fig:varying_rp}
\end{figure*}

\begin{figure*}
\begin{tabular}{ccc}
\includegraphics[width=3.in]{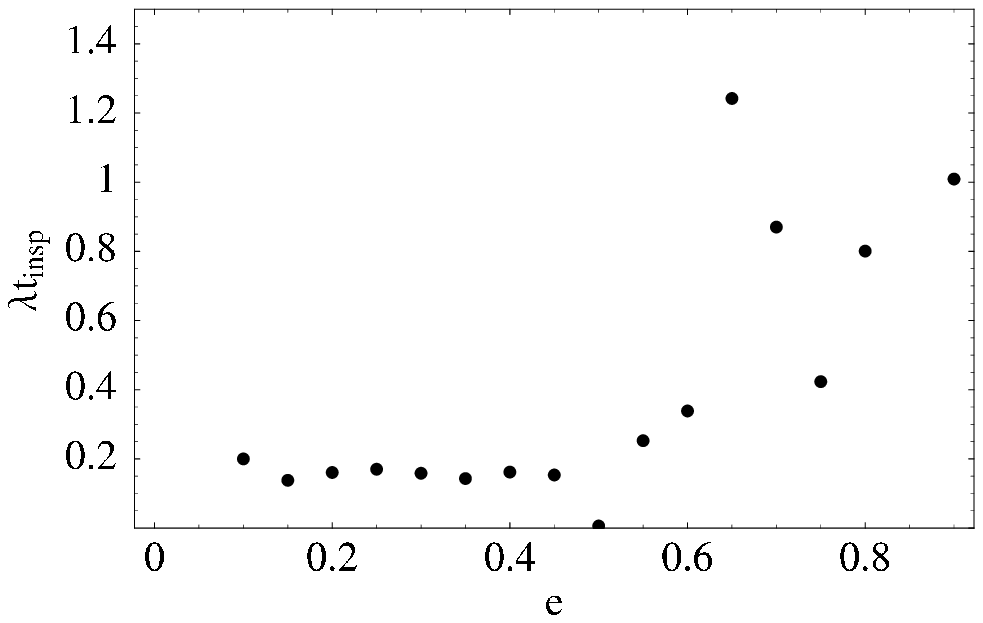} & \hspace{0.5in}
	& \includegraphics[width=3.in]{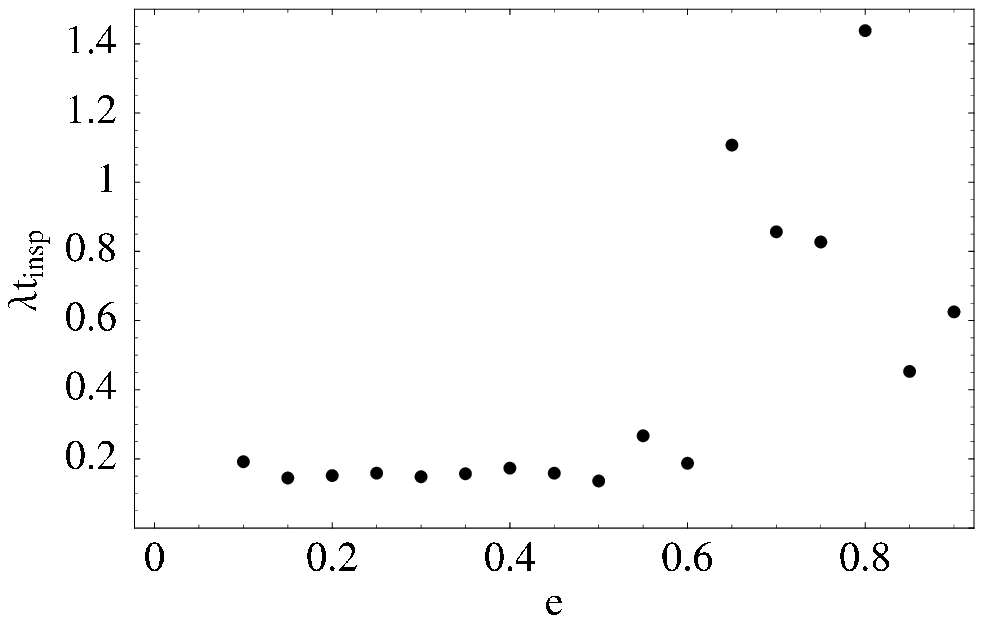}\\
(a) & & (b)\medskip\\
\end{tabular}
\caption{The dimensionless Lyapunov exponent $\lambda t_\mathrm{insp}$ as
a function of orbital eccentricity 
for (a) all terms through 2PN and (b) all terms
through 3PN. Only eccentricities greater than around 0.6 show any evidence of
chaos.
 The initial spins are the same as those in
Fig.~\ref{fig:eccentric_lyapunov_plot}, with the other initial conditions
fixed by the parameterization method described in
Sec.~\ref{sec:eccentric}.}
\label{fig:varying_e}
\end{figure*}

\begin{figure*}
\begin{tabular}{ccc}
\includegraphics[width=3.in]{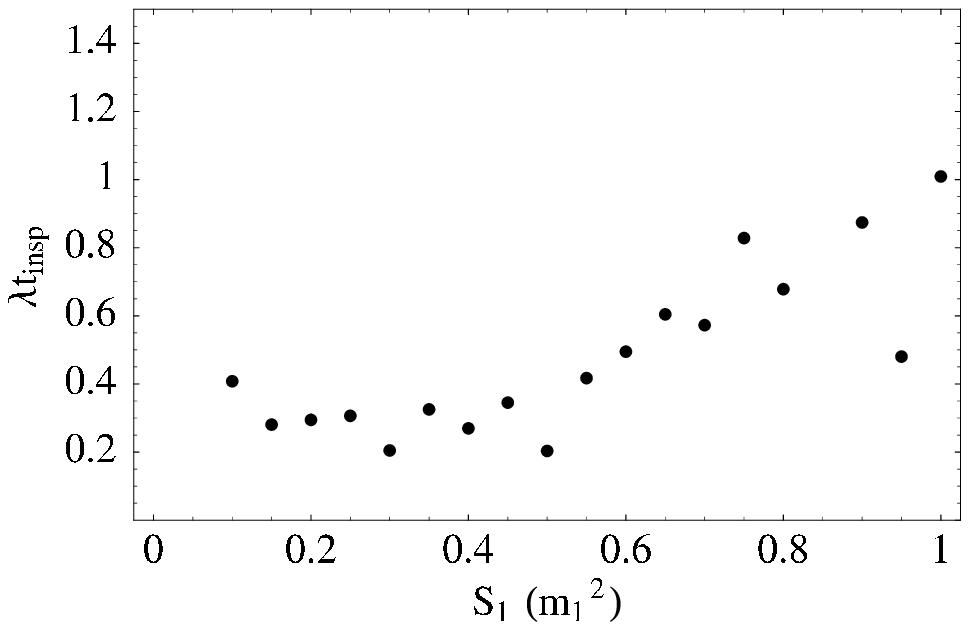} & \hspace{0.5in}
	& \includegraphics[width=3.in]{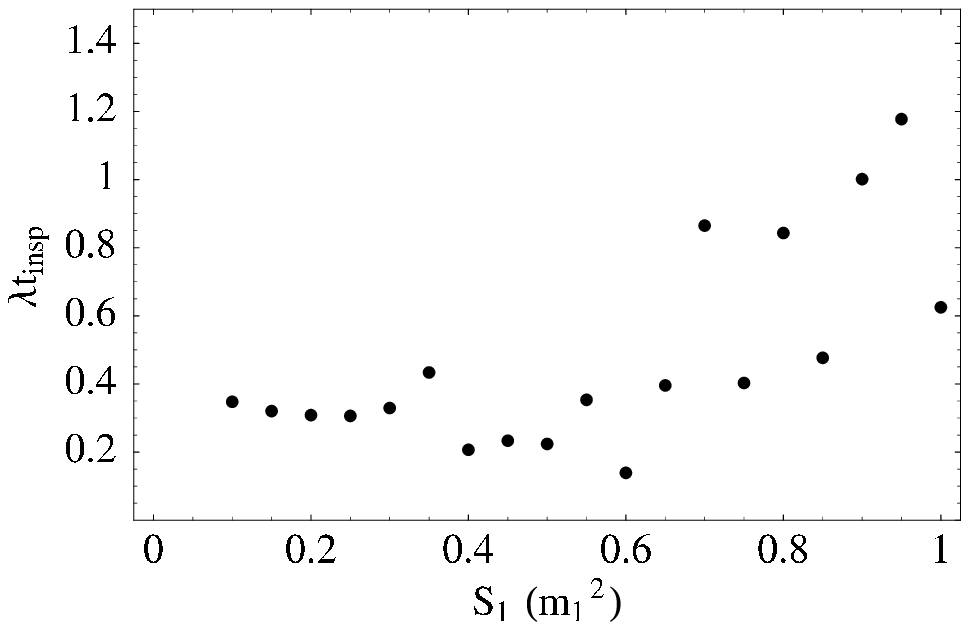}\\
(a) & & (b)\medskip\\
\end{tabular}
\caption{The dimensionless Lyapunov exponent $\lambda t_\mathrm{insp}$ as
a function of the first body's spin 
for (a) all terms through 2PN and (b) all terms
through 3PN. 
The initial spins are the same as those in
Fig.~\ref{fig:eccentric_lyapunov_plot}, with the other initial conditions
fixed by the parameterization method described in Sec.~\ref{sec:eccentric}. 
The strength of the chaos decreases as the spin decreases, becoming
indistinguishable from zero below around $S_1 = 0.5\,m_1^2$.}
\label{fig:varying_S}
\end{figure*}

In order to give a sense of the relative importance of various orbital and
spin parameters for the presence of chaos, we take the system shown in
Fig.~\ref{fig:eccentric_lyapunov_plot} and vary several parameters
independently. In Fig.~\ref{fig:varying_rp} we show the effect on the
dimensionless Lyapunov exponent $\lambda t_\mathrm{insp}$ of varying the
pericenter. Since the various spin terms (which make chaos possible) are
decreasing functions of $r$, we might expect that the chaos is weaker or
non-existent as the spin terms get smaller, and indeed this is the case: for
the model system we consider here, there is no chaos for $r_p > 6\,M$. 
Decreasing the eccentricity has a similar effect (Fig.~\ref{fig:varying_e});
more highly eccentric orbits are more likely to be chaotic, probably because
the larger velocities lead to larger values of the nonlinear
velocity-dependent terms in the Hamiltonian. Finally, varying the value of
the spin parameter for one of the bodies (Fig.~\ref{fig:varying_S}) produces
the expected result: as the spin decreases, the chaos is generally suppressed
or disappears altogether. From
Figs.~\ref{fig:varying_rp}--\ref{fig:varying_S}, it is clear that adding the
third post-Newtonian term has only minor effects in general on the presence
of chaos.

\subsubsection{A survey of eccentric orbits}

We undertake here a survey of eccentric orbits in an effort to understand the
prevalence of chaos in these systems. Unfortunately, in contrast to the
quasicircular case (Sec.~\ref{sec:quasicircular_survey}), any survey of
eccentric orbits is limited by the large number of parameters, which makes a
comprehensive survey impractical. Nevertheless, we have examined thousands
of initial conditions for a variety of masses and eccentricities, with
special attention paid to systems that are realistic sources of gravitational
radiation for ground-based detectors.

We consider binaries with masses of $(6+3) M_\odot$, $(6+4) M_\odot$,
and $(12+3) M_\odot$. Our choices for the eccentricities and pericenters are
then guided by astrophysical considerations~\cite{MH,LW}: eccentricities are
not larger than $ \sim 0.33$, and we choose the orbital frequency at
pericenter such that the corresponding (Newtonian) gravitational-wave
frequencies lie in the frequency band of ground-based interferometers  (for
low eccentricities GW radiation is emitted mostly at one, two, and three
times the  orbital frequency).  The specific values we consider here, which
satisfy the conditions above, are $e = 0.01, 0.2, 0.33$ and $f_\mathrm{orb} =
13.3, 20, 40, 50, 100$ Hz.  [Note that at eccentricity $ \sim 0.22$ ($ \sim
0.33$) the amplitudes of the Newtonian  gravity-wave signal for the first and
third harmonics are $17\%$ ($28\%$) and $40\%$ ($60\%$)  relative to the
second harmonic~\cite{MBM}.]  The pericenter is then obtained from the
formula~\cite{MartelPoisson1999}
\begin{equation}
\label{eq:pericenter}
r_p = \frac{1-e}{(2\pi M f_{\rm orb})^{2/3}}.
\end{equation}
With the choices made for $e$ and $f_\mathrm{orb}$, Eq.~(\ref{eq:pericenter})
takes values  between $8\,M$ and $30\,M$. Thus, the PN expansion is valid for
the two compact bodies even at the point of closest approach.

For each value of the binary masses and for each pair ($e$, $f_\mathrm{orb}$)
we produce  500 randomly oriented maximal initial spins and calculate the
largest Lyapunov exponent as in Sec.~\ref{sec:quasicircular_survey}, using a
final  integration time of ten times the inspiral time (as calculated
from~\cite{Peters1964}).  [For each random orientation we first fix the
spin, and then use the method described in Sec.~\ref{sec:eccentric} to find
the initial conditions; in particular, the PN Hamiltonian in
Eq.~(\ref{eq:HeqE}) always includes the spin contributions.] In all cases
considered, we find no evidence of chaos---all the Lyapunov exponents  are
consistent with zero. 

\section{Conclusions}

The dynamics of binary black holes, as modeled by the post-Newtonian equations,
are significantly affected by the presence of spin. In particular, the
addition of spin terms to the post-Newtonian equations of motion leads to
significant changes in the orbital geometry and dynamical behavior of the
solutions. The effects of the spin terms are particularly clear on
quasicircular orbits, where the interaction terms quadratic in the spin cause
deviations from perfectly spherical orbits. 

We find that, for quasicircular orbits, the presence of the interaction
terms quadratic in the spins can lead to chaotic solutions, as indicated by
positive Lyapunov exponents. These exponents come in $\pm\lambda$ pairs, 
a reflection of the Hamiltonian nature of the dynamics. We measure the
strength of the chaos by comparing the $e$-folding timescale for chaotic
behavior (the inverse of the Lyapunov exponent) with the inspiral timescale.
We find especially strong chaos for high-frequency/low-radius orbits and high
spins. However, in those cases the black holes are so close to each other
that spin-spin induced oscillations in the radial separation are rather
important, and the quasicircular initial conditions used in this paper 
may not correspond to widely separated quasicircular orbits evolved adiabatically to low
separations under gravitational radiation reaction. It would therefore be
preferable to set initial conditions when the black holes are rather far
apart (so that spin effects are negligible) and then evolve the system
including radiation-reaction effects. The latter will be soon
available~\cite{BCD} in the Hamiltonian framework, including spin couplings,
and we plan to include them and investigate the presence of chaos in the
near future. 

We build a survey of eccentric orbits which we believe is representative  of
binaries which can emit GWs in the LIGO/VIRGO frequency band, have
eccentricities justified by astrophysical considerations, and whose dynamics
can be safely described by the PN expansion. For this survey we do not find
any chaos. We find chaotic solutions only for rather eccentric orbits ($e
\sim 0.9$) with very low pericenters, which are not astrophysically motivated
and for which higher order PN terms, especially higher-order spin couplings,
should be consistently added in the ADM Hamiltonian. 

Since we find that chaotic behavior is due to spin-spin couplings and does
not seem to be much affected by the non-spinning PN dynamics, we suspect
that our results will not change qualitatively if the non-spinning ADM
Hamiltonian were replaced by the Schwarzschild deformed effective-one-body 
Hamiltonian~\cite{BD1,BD2} (which is a re-summation of the ADM Hamiltonian).
By contrast, we might expect differences if the Kerr-deformed
effective-one-body Hamiltonian~\cite{Damour2001} were used instead of the
spinning ADM Hamiltonian.

In conclusion, considering our surveys of quasicircular and eccentric orbits
together, we find no chaos in any system for orbits that are of astrophysical
interest for ground-based interferometers and which clearly satisfy the
approximations required for the equations of motion to be physically valid
at the post-Newtonian order considered. 

\acknowledgments We wish to thank Yanbei Chen, Thibault Damour, and Sterl
Phinney for useful interactions and Achamveedu Gopakumar for stimulating
comments.  M.~D.~Hartl acknowledges the support of NASA grant NAG5-10707.

\appendix*
\section{Numerical Implementation}
\label{sec:appendix}

Our primary implementation of the post-Newtonian equations of motion is a
collection of \emph{Mathematica} packages, relying on the native
\texttt{NDSolve} function to effect numerical integrations. We implement the
equations in a standard way to eliminate the mass variables, measuring
lengths and times in terms of the total mass~$M$  and momenta in terms of the
reduced mass~$\mu$.  In these units, consistency then forces the angular
momenta to be measured in terms of $\mu M$. 

We perform various tests to
check the validity of the equations of motion and the numerical
integrations.  Most important, we check that the energy, angular momentum,
and spin magnitudes are all conserved by the integrations. These quantities
are useful since they tend to be sensitive to mistakes in the equations of
motion; we verify in all cases that the constants of the motion are conserved
at a level consistent with the default accuracy goal (typically $10^{-10}$),
which gives us confidence that the equations are correct. 
We also check that orbital angular momentum and
$\mathbf{L}\cdot\mathbf{S}_\mathrm{eff}$ are conserved through 
spin-orbit coupling, as discussed in Sec.~\ref{sec:conserved}.  Furthermore,
we verify that our implementation of the equations reproduces the Keplerian
orbits of Newtonian gravitation (only Newtonian terms turned on), Lense-Thirring precession
(extreme mass-ratio $m_1\gg m_2$ limit with Newtonian and SO terms turned on), and
classical quadrupole precession (Newtonian and $S_1S_2$ turned on).

A second implementation of the equations of motion uses a \emph{Mathematica}
interface for a C++ numerical integrator. We use \emph{Mathematica} to
generate the required derivative expressions directly from the Hamiltonian,
using the built-in \texttt{CForm} function to convert to C++ source code, so
that the native \emph{Mathematica} and C++ integrators automatically
agree.\footnote{The additional use of the freely available package
\texttt{Optimize.m}~\cite{Optimize} gives a factor of~5 increase in speed for
the case at hand.}  As in the case of the pure \emph{Mathematica} integrator,
we check the conservation of all the relevant quantities, in this case using
a Bulirsch-Stoer integrator with an error goal of $10^{-10}$ (the variable
\texttt{eps} from Numerical Recipes~\cite{NumRec}).  We also check several
orbits with a Runge-Kutta integrator with an error goal of $10^{-9}$; the
resulting agreement with the Bulirsch-Stoer integrations verifies that our
results are not specific to the choice of integration algorithm.

As noted in Sec.~\ref{sec:lyapunov}, for the calculation of Lyapunov
exponents we use the techniques and routines described in \cite{Hartl_2002_1}
and especially Chapter~5 of \cite{Hartl_thesis}.  In short, we use the
deviation vector method for speed, but check select results using the slower
but more robust Jacobian method.  We also use the Jacobian diagnostic
described in \cite{Hartl_2002_1} and \cite{Hartl_thesis} to verify the
correctness of the (rather complicated) Jacobian matrix (which in fact is
generated by \emph{Mathematica} directly from the equations of motion).  As a
result, we are confident that our values for the Lyapunov exponents
faithfully describe the true dynamics of the system.

\bibliography{bibliography}

\end{document}